\documentclass[superscriptaddress,showpacs,amssymb,10pt,reprint,aps,prd,longbibliography,nofootinbib,floatfix]{revtex4-1}

\usepackage{graphicx,epsfig,amssymb} 
\usepackage{amsmath,amsfonts, times}
\usepackage{bm} 
\usepackage{epstopdf}
\usepackage[linktocpage,colorlinks]{hyperref}
\usepackage[caption=false]{subfig}
\usepackage[usenames]{color}     
\usepackage{natbib}
\usepackage{subfig}
\usepackage[utf8x]{inputenc}
\usepackage{soul}
\usepackage[dvipsnames]{xcolor}
\usepackage[normalem]{ulem}

\newcommand\numberthis{\addtocounter{equation}{1}\tag{\theequation}}

\begin{document}
\title{\large Singular space-times with bounded algebraic curvature scalars}

\author{Renan B. Magalh\~aes}
\email{renan.magalhaes@icen.ufpa.br}
\affiliation{Programa de P\'os-Gradua\c{c}\~{a}o em F\'{\i}sica, Universidade 
	Federal do Par\'a, 66075-110, Bel\'em, Par\'a, Brazil.}

\author{Gabriel P. Ribeiro}
\email{gabriel.ribeiro@icen.ufpa.br}
\affiliation{Programa de P\'os-Gradua\c{c}\~{a}o em F\'{\i}sica, Universidade 
	Federal do Par\'a, 66075-110, Bel\'em, Par\'a, Brazil.}

\author{Haroldo C. D. Lima Junior}
\email{haroldo.ufpa@gmail.com}
\affiliation{Programa de P\'os-Gradua\c{c}\~{a}o em F\'{\i}sica, Universidade 
	Federal do Par\'a, 66075-110, Bel\'em, Par\'a, Brazil.}

\author{Gonzalo J. Olmo}
	\email{gonzalo.olmo@uv.es}
	\affiliation{Departamento de F{\'i}sica Te{\'o}rica and \textit{IFIC}, Centro Mixto Universitat de Val\`encia - \textit{CSIC}. Universitat de Val\`encia, Burjassot-46100, Val\`encia, Spain.}
	\affiliation{Universidade Federal do Cear\'a (UFC), Departamento de F\'isica,\\ Campus do Pici, Fortaleza - CE, C.P. 6030, 60455-760 - Brazil.}

\author{Lu\'is C. B. Crispino}
\email{crispino@ufpa.br}
\affiliation{Programa de P\'os-Gradua\c{c}\~{a}o em F\'{\i}sica, Universidade 
	Federal do Par\'a, 66075-110, Bel\'em, Par\'a, Brazil.}

\begin{abstract}
We show that the absence of unbounded algebraic curvature invariants constructed from polynomials of the Riemann tensor cannot guarantee the absence of strong singularities. As a consequence, it is not sufficient to rely solely on the analysis of such scalars to assess the regularity of a given space-time. This conclusion follows from the analysis of incomplete geodesics within the internal region of asymmetric wormholes supported by scalar matter which arise in two distinct metric-affine gravity theories. These wormholes have bounded algebraic curvature scalars everywhere, which highlights that their finiteness does not prevent the emergence of pathologies (singularities) in the geodesic structure of space-time. 
By analyzing the tidal forces in the internal wormhole region, we find that the angular components are unbounded along incomplete radial time-like geodesics. The strength of the singularity is determined by the evolution of Jacobi fields along such geodesics, finding that it is of strong type, as volume elements are torn apart as the singularity is approached. Lastly, and for completeness, we consider the wormhole of the quadratic Palatini theory and present an analysis of the tidal forces in the entire space-time.
\end{abstract}

\date{\today}

\maketitle

\section{Introduction}\label{sec:int}
According to general relativity (GR), gravitational collapse can lead to the formation of different compact objects such as neutron stars and black holes. There is also theoretical evidence that, depending on the distribution of matter, this final stage may lead to the formation of a naked singularity (see, for instance, \cite{Joshi:1993,Mosani:2020}), in contrast with the widely accepted belief that singularities should always be hidden behind event horizons~\cite{Penrose:1969}. If naked singularities could indeed exist, they would potentially display distinctive phenomenology, such as their shadow image~\cite{Gyulchev:2019}, which would set them apart from other types of compact objects. Consequently, understanding the nature and properties of space-time singularities is crucial for both phenomenological and theoretical discussions and interpretations.

In a first approximation to the problem, space-time singularities could be described as locations \textit{in} space-time where the laws of Physics break down because the curvature or other physical quantities become infinite. This intuitive notion has undergone refinement and revision over the years. 
Remarkable advances in this sense were made in the 1960s with the singularity theorems~\cite{Penrose:1965,Senovilla:2014gza,Witten:2020}. 
According to these theorems, the classification of whether a space-time is singular (or not) is not based on the presence of divergent quantities such as curvature scalars or energy densities. Instead, it relies on the existence of incomplete curves in those space-times~\cite{Geroch:1968}. For instance, the story of observers following time-like incomplete geodesics necessarily ends (or starts) at a finite value of their proper time, implying that they can be destroyed (or created). Something similar applies to light-like curves, which transport information. Incomplete geodesics, therefore, would lead to the creation/destruction of information and to our inability to perform physical measurements (when observers are destroyed or unable to exist). 

In the classical framework of GR, the singularity theorems do not offer information on the attributes, positions, or quantities of singularities that may exist within a given space-time. These theorems guarantee that, provided reasonable energy conditions are met, the gravitational collapse of matter distributions will inevitably result in the formation of space-time singularities if trapped surfaces are produced~\cite{Hawking:1973}.
One would like, however, to have more information about the nature and signatures of possible singularities existing in a given space-time. For example, one may be interested in whether physical or curvature quantities behave well when approaching these singularities along time-like or null trajectories. Depending on this behavior, space-time singularities can be regarded as curvature or non-curvature singularities, respectively~\cite{Ellis:1979}.

Curvature singularities are commonly associated with unbounded scalar polynomials of the curvature tensor, such as the Ricci or Kretschmann scalars, along an incomplete causal curve. However, even if all curvature invariants\footnote{The set of all polynomial curvature invariants includes curvature invariants of any order. A curvature invariant of order $k$ consists of a scalar constructed as contractions of the metric, the Riemann tensor and its covariant derivatives of order $\leq k$. Throughout this paper, we simply call algebraic curvature scalars the zeroth-order polynomial curvature invariants.} are bounded everywhere, a different type of curvature singularity can occur~\cite{Hirschmann:2004}. These \textit{non-scalar} curvature singularities arise from the divergence of any component of the curvature tensor in a parallelly propagated basis attached to an inextensible causal geodesic. For example, one can investigate the finiteness of the tidal forces along incomplete causal curves. If at least one component of the tidal tensor is unbounded, then the space-time is said to have a parallelly propagated curvature singularity~\cite{Clarke:1975}. 

Studying space-times that have bounded curvature scalars but still have inextensible curves can provide valuable insights into the physical mechanisms behind space-time singularities. 
 However, finding examples of such objects in GR is challenging since most known naked singularity solutions exhibit scalar curvature singularities~\cite{Janis:1968,Wyman:1981}. Extensions of GR may help in this regard. In fact, it has been shown that free scalar fields coupled with some Ricci-Based Gravity theories (RBGs) in the Palatini framework may give rise to geodesically incomplete asymmetric wormhole solutions in which algebraic curvature invariants are bounded everywhere~\cite{AORG:2017,MCO:2022}. Despite the behavior of their curvature scalars, observers moving towards the interior of the wormhole reach the asymptotic internal region in a finite proper time, which is a clear indication of the existence of physical pathologies in such space-times. A detailed analysis of what goes wrong in such geometries is not yet available, and that is precisely what motivates this work.

These singular asymmetric wormholes arise in two different gravity theories coupled to free scalar fields, namely, i) Palatini $f({\cal R},{\cal R}_{(\mu\nu)}{\cal R}^{(\mu\nu)})={\cal R}+a{\cal R}^2+b{\cal R}_{(\mu\nu)}{\cal R}^{(\mu\nu)}$ gravity coupled to a non-linear matter Lagrangian~\cite{MCO:2022}, and ii) the Eddington-inspired Born-Infeld (EiBI) gravity model coupled to a canonical scalar matter Lagrangian~\cite{AORG:2017}. Surprisingly, despite the different physical settings (different gravity theories with different scalar Lagrangians), the asymptotic interior region of the resulting asymmetric wormholes can be approximated by the same line element (up to model-dependent constants)~\cite{MCO:2022}. 
Here we investigate in detail the geodesic structure of this asymptotic region, discussing the non-scalar nature of the singularity, whose presence is manifest from the existence of radial time-like and null incomplete geodesics. 

The analysis of tidal forces around compact objects is important for both observational and theoretical reasons. Tidal forces play a significant role in the so-called tidal disruption radius, also known as the Roche radius. The Roche radius is the radial distance to the center of a compact object such that a star is tidally disrupted. The tidal disruption of stars by black holes has been conjectured to be the cause of the flares observed in different branches of the electromagnetic spectrum, such as optical~\cite{TDE1}, ultraviolet~\cite{TDE2} and X-ray~\cite{TDE3}. It is important to investigate the tidal forces for different compact objects, such as naked singularities and wormholes, as they may give rise to signatures that deviate from those predicted by black holes. In this work, we perform an analysis of the tidal forces and Jacobi fields in order to characterize the type and strength of the singularity in the interior of the family of wormholes considered here. Our analysis shows that the interior region has a strong parallelly propagated curvature singularity. As a result, anyone moving towards the asymptotic region would be torn apart due to the emergence of infinite tidal forces.

The content of this paper is organized as follows. In Sec.~\ref{sec:nrwhRBGs} we give an overview of the asymmetric wormholes that arise in the Ricci-Based Gravity theories \textit{\`a la} Palatini mentioned above. In Sec.~\ref{sec:appx} we discuss the approximations of the interior region of these wormholes, investigating their curvature scalars and geodesic structure. In Sec.~\ref{sec:pp} we perform an analysis of the tidal forces to characterize the type of singularity in the interior of the wormhole. Moreover, we also study the Jacobi fields approaching the singularity and discuss its strength. For completeness, in Sec.~\ref{sec:tidal_QP}, we make an analysis of the tidal forces in the whole space-time of one of the singular asymmetric wormholes, namely the one associated to the quadratic Palatini gravity. Finally, we summarize our results and discuss some perspectives in Sec.~\ref{sec:final}.

\section{Two gravity models with scalar matter}\label{sec:nrwhRBGs}
Extensions of GR in the Palatini formalism, such as $f({\cal R})$ and $f({\cal R},{\cal R}_{(\mu\nu)}{\cal R}^{(\mu\nu)})$ gravities, have been studied in the literature, allowing, for instance, to produce modified dynamics without the introduction of new degrees of freedom~\cite{Olmo:2011}. Remarkably, in the cases of $f({\cal R})$ and $f({\cal R},{\cal R}_{(\mu\nu)}{\cal R}^{(\mu\nu)})$ gravities, the field equations in vacuum reduce to those of General Relativity with an effective cosmological constant that depends on the model~\cite{Ferraris:1992,Borowiec:1996}. Other sort of Palatini gravities, with other contractions of the Ricci tensor, have also been considered in the literature, such as the inspired by Born-Infeld theories~\cite{Vollick:2003qp,BeltranJimenez:2017,Afonso:2021aho}. All these theories can be cast as members of a family of extensions of GR, the so-called RBGs, which can be described by an action of the form~\cite{Gonzalo:2022}
\begin{align*}
\label{eq:action_RBGs}
S_{RBG}(g_{\mu\nu},\Gamma^{\alpha}_{\beta\gamma},\psi_m)=&\dfrac{1}{2\kappa^2}\int d^4x\sqrt{-g}f(g^{\mu\alpha}{\cal R}_{(\alpha \nu)})\\&+S_m(g_{\mu\nu},\psi_m),\numberthis
\end{align*}
where $\kappa$ is a coupling constant in suitable units; $g_{\mu\nu}$ and $g$ denote the components of the metric tensor and its determinant, respectively; $\Gamma^{\alpha}_{\beta\gamma}$ is the affine connection, which is independent of the space-time metric since we are in the Palatini picture, that define the Riemann tensor ${\cal R}^\alpha{_{\beta\mu\nu}}$; and $\psi_m$ denotes any sort of matter field present in the space-time. From now on, let us denote ${\cal R}\equiv g^{\mu\nu}{\cal R_{(\mu\nu)}}$ and ${\cal Q}\equiv {\cal R_{(\mu\nu)}}{\cal R^{(\mu\nu)}}$. By varying the action~\eqref{eq:action_RBGs} with respect to the metric and affine connection, independently, one obtains the field equations associated to these two fields. The source of the metric field equation is the so-called energy-momentum tensor, $T_{\mu\nu} \equiv (-2/\sqrt{-g})\delta S_m/\delta g^{\mu\nu}$, while the source of the connection equation, the so-called hypermomentum tensor, is identically zero, since we are not assuming couplings between matter fields and the affine connection~\cite{Afonso:2017}. Since in action~\eqref{eq:action_RBGs} only the symmetric part of the Ricci tensor appears, it is projective invariant, thus ensuring the absence of ghosts in the theories~\cite{BeltranJimenez:2019,BeltranJimenez:2020}.

By introducing an effective metric, $h_{\mu\nu}$, with determinant $h$, related to the space-time metric by $h_{\mu\nu} = g_{\mu\alpha}\Omega^{\alpha}{_\nu}$, where $\Omega^{\alpha}{_\nu}$ is the so-called deformation matrix, one obtains the metric and connection field equations in the Einstein-frame representation given, respectively, by~\cite{Gonzalo:2022}
\begin{align}
\label{eq:metric_RBG}&G^{\mu}{_\nu}(h) = \dfrac{\kappa^2}{\sqrt{|\Omega|}}\left[T^{\mu}{_\nu}-\dfrac{1}{2\kappa^2}(f+\kappa^2 T)\delta^{\mu}{_\nu}\right],\\
\label{eq:connection_RBG}&\nabla_\alpha(\sqrt{-h}h^{\mu\nu})= 0,
\end{align}
where $G^{\mu}{_\nu}(h) = h^{\mu\alpha}(R_{\alpha\nu}(h)+h_{\alpha\nu}R(h)/2)$ is the Einstein tensor of the auxiliary metric $h_{\mu\nu}$, $|\Omega|$ is the determinant of $\Omega^{\alpha}{_\nu}$ (from now on, vertical bars also denote determinant) and $T$ is the trace of the energy-momentum tensor.

Coupled to the RBG theories, let us consider a scalar field distribution with matter action given by
\begin{equation}
\label{eq:action_matter}
S_m(X,\phi) = -\dfrac{1}{2}\int d^4x \sqrt{-g}P(X,\phi),
\end{equation}
where $P(X,\phi)$ is some function of the kinetic term $X\equiv g^{\mu\nu}\partial_\mu \phi \partial_\nu \phi$ and the scalar field $\phi$. The energy-momentum tensor related to the scalar action~\eqref{eq:action_matter} reads
\begin{equation}
\label{eq:scalar_Tmunu}
T^{\mu}{_{\nu}} = P_X X^{\mu}{_\nu} - \dfrac{P(X,\phi)}{2}\delta^{\mu}{_\nu},
\end{equation} 
where $P_X\equiv \partial P/\partial X$ and $X^{\mu}{_\nu}\equiv g^{\mu\alpha}\partial_\alpha \phi \partial_\nu \phi$. The evolution equation of the scalar field comes from the variation of the action~\eqref{eq:action_matter} with respect to the scalar field, namely
\begin{equation}
\label{eq:scalar_ev}
\dfrac{1}{\sqrt{-g}}\partial_{\mu}(P_X \sqrt{-g} g^{\mu\alpha}\partial_\alpha\phi)-\dfrac{P_\phi}{2}=0,
\end{equation}
where $P_\phi\equiv \partial P/\partial \phi$.

In the next two sections, we discuss wormhole solutions supported by scalar fields in two different RBG theories coupled with distinct matter actions but which, nonetheless, end up producing a similar internal region.

\subsection{Quadratic Palatini $f({\cal R},{\cal Q})$ gravity}\label{sec:f(R,Q)}
It was shown in Ref.~\cite{AORG:2018}, that the space of solutions of GR coupled to a scalar field may be mapped into a wide family of Palatini RBGs. The explicit mappings to Palatini $f({\cal R})$ and EiBI gravities were implemented in Refs.~\cite{AOORG:2018,AOORG:2019}, whereas the mapping to Palatini $f({\cal R},{\cal Q})$ gravity was worked out in Ref.~\cite{MCO:2022}. In those papers one finds that starting with a static, spherically symmetric solution of GR canonically coupled to a free scalar field, one can obtain solutions in RBGs coupled to a non-linear matter Lagrangian that inherits the nonlinearities of the gravitational sector.

The key idea behind the mapping procedure relies on a suitable reinterpretation of Eq.~\eqref{eq:metric_RBG}. The right-hand side of this equation can be visualized as an effective energy-momentum tensor, such that Eq.~\eqref{eq:metric_RBG} may be cast as the  equations of GR for the auxiliary metric $h_{\mu\nu}$ where the matter source is a non-linear scalar field minimally coupled to $h_{\mu\nu}$. By relating the energy-momentum tensor of the original RBG theory (where the scalar field was minimally coupled to $g_{\mu\nu}$) to the effective one in the {\it Einstein frame}, we can derive the mapping equations and the deformation matrix that connects the space-time and the auxiliary metrics. Once we have the mapping equations and the deformation matrix, we can solve the Einstein's equations for the auxiliary metric, and use this solution as a seed to obtain a solution in the corresponding RBG model.

For concreteness, let us now consider the quadratic Palatini (QP) gravity described by the action
\begin{equation}
\label{eq:action_quadr}
S_{QP} = \dfrac{1}{2\kappa^2}\int d^4x \sqrt{-g} \left({\cal R}+a{\cal R}^2  +b {\cal Q}\right),
\end{equation}
where $a$ and $b$ are constants with dimensions of length squared. It was shown in Ref.~\cite{MCO:2022} that one can obtain solutions of the QP model~\eqref{eq:action_quadr} using the mapping method by coupling the action~\eqref{eq:action_quadr}
to a nonlinear scalar field with action 
\begin{equation}
\label{eq:matter_quad}
S_m(X,\phi) = -\dfrac{1}{2}\int d^4x\sqrt{-g}\left(X+(a+b)\kappa^2 X^2\right).
\end{equation}
The mapping method establishes that the space of solutions of the theory (\ref{eq:action_quadr}) and (\ref{eq:matter_quad}) is in correspondence with the space of solutions of GR minimally coupled to a free scalar field of the form 
\begin{equation}
\label{eq:canonical_coupling_GR}
\tilde{S}_{m}(Z) = -\dfrac{1}{2}\int d^4 \sqrt{-h} Z,
\end{equation}
where $Z\equiv h^{\mu\nu}\partial_\mu \phi \partial_\nu \phi$ is the kinetic term of the scalar field associated with the auxiliary metric. When one considers static, spherically symmetric solutions, the GR problem leads to the Wyman solution~\cite{Wyman:1981} (also discovered before by other authors~\cite{Janis:1968}), which describes a naked singularity. By considering the scalar field as a radial coordinate, one can cast the Wyman solution in the form
\begin{equation}
\label{eq:wyman}
ds^2_{GR} = -e^{\nu}\,dt^2 + \dfrac{e^{\nu}}{W^4}\,dy^2 + \dfrac{1}{W^2}\,(d\theta^2 + \sin^2\theta d\varphi^2),
\end{equation}
such that the scalar field satisfies the simple Klein-Gordon equation 
\begin{equation}
\label{eq:scalar_field_equation}
\phi_{yy}=0,
\end{equation}
that has $\phi=\zeta y$ as a solution, where $\zeta$ is a constant related with the amplitude of the scalar field. From the requirement of the asymptotic flatness of the Wyman solution, the functions $\nu$ and $W$ are given by~\cite{Wyman:1981}
\begin{align}
\nu &= -2M\,y,\\
W&= \dfrac{e^{-M\,y}\sinh(\eta\,y)}{\eta}, \label{eq:nuW}
\end{align}
where $M$ is the asymptotic ADM mass of the solution and $\eta \equiv \sqrt{M^2 + \mu^2/2}$, with $\mu^2\equiv \kappa^2\zeta^2$.

From the mapping, one finds that solutions from GR coupled to the action~\eqref{eq:canonical_coupling_GR} are \textit{disformally} related to solutions of Palatini $f({\cal R},{\cal Q})$ coupled to the action~\eqref{eq:matter_quad} via the deformation matrix~\cite{MCO:2022}
\begin{equation}
\label{eq:def_matrix_quad}
\Omega^{\mu}{_\nu} = \sqrt{f_{\cal R}F_X}\left(\delta^{\mu}{_\nu} - \dfrac{2\kappa^2 f_{\cal Q}}{F_X}{X^\mu}_{\nu}\right),
\end{equation}
where $F_X=f_{\cal R}+2\kappa^2 X f_{\cal Q}$, with $f_{\cal R}$ and $f_{\cal Q}$ taking the explicit form
\begin{align}
f_{\cal R} &= 1+2a\kappa^2 X\\
f_{\cal Q} &= b \ ,
\end{align}
respectively. From the above, it follows that the relation between $Z=h^{\mu\nu}\partial_\mu \phi \partial_\nu \phi$ and $X=g^{\mu\nu}\partial_\mu \phi \partial_\nu \phi$ is given by
\begin{equation}
\label{eq:relation_Z_X}
Z = X\sqrt{\dfrac{F_X}{f_{\cal{R}}^3}}.
\end{equation}
In order to find $X(Z)$, one must solve a cubic polynomial equation that generates three roots $X_i(Z)$ ($i=1$, 2 or 3), which must be carefully combined in order to build the right solutions on the physically relevant interval. Depending on the parameters $a$ and $b$ considered, taking into account the limit $X\approx Z$ for small values of $Z$, the desired solutions can be obtained by requiring the continuity and differentiability of the resulting curve. 

Finally, the counterpart of the GR solution~\eqref{eq:wyman} obtained via the mapping method in the QP theory~\eqref{eq:action_quadr}  is given by
\begin{align*}
ds^2 &= -\dfrac{e^{\nu}}{\sqrt{f_{\cal R}F_X}}\,dt^2 + \left(\dfrac{e^{\nu}}{\sqrt{f_{\cal R}F_X}W^4}+\dfrac{2\mu^2b}{F_X}\right)\,dy^2 \\&+ \dfrac{1}{\sqrt{f_{\cal R}F_X}W^2}\,(d\theta^2 + \sin^2\theta d\varphi^2). \numberthis \label{eq:ds_quadr}
\end{align*}
In the limit that the QP model~\eqref{eq:action_quadr} goes to the Starobinsky model, that is $b\to 0$, the line element~\eqref{eq:ds_quadr} becomes conformally related to~\eqref{eq:wyman}, and one obtains the solutions found in Refs.~\cite{AOORG:2018,AOORG:2019}. Here we focus in the case $b\neq 0$, such that the relation between the metrics is disformal. The properties of the line element~\eqref{eq:ds_quadr} for the positive and negative branch of $a$ were investigated in Ref.~\cite{MCO:2022}. Here, in particular, we are interested in the negative branch of $a$, since it presents a throat-like structure (a regular minimum in $r^2(y)=1/(\sqrt{f_{\cal R}F_X}W^2)$) connecting two asymmetric universes. The external one is asymptotically flat, whereas the internal one describes a non-asymptotically flat space-time. A remarkable feature of this solution is that all its algebraic curvature invariants are bounded.

\subsection{Eddington-inspired Born-Infeld gravity}\label{sec:eibi}

We now consider static and spherically symmetric solutions of a free scalar field minimally coupled to the Palatini EiBI theory, whose gravitational sector is given by~\cite{Vollick:2003qp,BeltranJimenez:2017}
\begin{equation}
\label{eq:action_EiBI}
S_{EiBI}=\dfrac{1}{\epsilon \kappa^2}\int d^4x \left[-|g_{\mu\nu}+\epsilon {\cal R}_{(\mu\nu)}|-\lambda\sqrt{-g}\right] \ .
\end{equation}
Here $\epsilon$ is a parameter with dimensions of length squared. By taking a perturbative expansion in $\epsilon$ for the fields $|{\cal R}_{(\mu\nu)}|\ll 1/\epsilon$ on the action~\eqref{eq:action_EiBI}, one obtains the field equations of GR plus an effective cosmological constant, that is $\text{GR}+\Lambda_{\text{eff}}+{\cal O}(\epsilon)$. From that expansion in $\epsilon$, it reads that the parameter $\lambda$ is related to the effective cosmological constant as $\lambda=1+\epsilon\kappa^2\Lambda_{\text{eff}}$. A glance at action~\eqref{eq:action_EiBI} suggests the convenient redefinition $h_{\mu\nu} = g_{\mu\nu} + \epsilon {\cal R}_{(\mu\nu)}$. From this choice, the metric field equation becomes
\begin{equation}
\label{eq:metric_field_EiBI}
\sqrt{-h}h^{\mu\nu}-\sqrt{-g}\lambda g^{\mu\nu}=-\kappa^2\epsilon\sqrt{-g}T^{\mu\nu},
\end{equation}
while the connection field equation is given by Eq.~\eqref{eq:connection_RBG}, which implies that the affine connection is the Levi-Civita connection of the auxiliary metric  $h_{\mu\nu}$. From Eq.~\eqref{eq:metric_field_EiBI} one finds that the deformation matrix also has a disformal structure,
\begin{equation}
\label{eq:deformation_EiBI}
\sqrt{|\Omega|}(\Omega^{-1})^{\mu}{_\nu} = \lambda \delta^{\mu}{_\nu}-\epsilon\kappa^2T^{\mu}{_\nu}.
\end{equation}

Coupled to EiBI, let us consider the canonical scalar Lagrangian, that is $P(X,\phi)=X-2V(\phi)$, where $V(\phi)$ is the potential, for which the energy-momentum tensor and the dynamical equation of the scalar field, respectively, reduce to
\begin{align}
\label{eq:T_scalar_canonical}&T^{\mu}{_\nu} =  X^{\mu}{_\nu} - \dfrac{(X-2V(\phi))}{2}\delta^{\mu}{_\nu},\\
\label{eq:kg_equation_canonical}&\dfrac{1}{\sqrt{-g}}\partial_{\mu}(\sqrt{-g} g^{\mu\alpha}\partial_\alpha\phi)-\dfrac{dV}{d\phi}=0.
\end{align}
One can follow Ref.~\cite{AORG:2017} in order to find scalar configurations generated by a self-gravitating free scalar field ($V(\phi)=0$), and consider the following line element of a generic spherically symmetric space-time,
\begin{equation}
\label{eq:ss_ds_EiBI}
ds^2=-A(x)+\dfrac{1}{B(x)}dx^2+r^2(x)(d\theta^2+\sin^2\theta d\varphi^2).
\end{equation}
By substituting the line element~\eqref{eq:ss_ds_EiBI} in the scalar field equation, one can integrate it to obtain $r^2\sqrt{AB}\phi_x=C$, where $\phi_x=d\phi/dx$ and $C$ is an integration constant. From Eqs.~\eqref{eq:deformation_EiBI} and~\eqref{eq:T_scalar_canonical}, together with the line element~\eqref{eq:ss_ds_EiBI}, we see that a suitable \textit{ansatz} for the deformation matrix is
\begin{align*}
\Omega^{\mu}{_\nu} &=\text{diag}(\Omega^{t}{_t},\Omega^{\theta}{_\theta},\Omega^{\varphi}{_\varphi},\Omega^{x}{_x})\\
&=\text{diag}(\Omega_+,\Omega_+,\Omega_+,\Omega_-), \numberthis
\end{align*}
where $\Omega_+ = (\lambda^2-X_{\epsilon}^2)^{1/2}$, $\Omega_- = (\lambda+X_{\epsilon})^{3/2}(\lambda-X_{\epsilon})^{-1/2}$
and $X_\epsilon$ is defined by $X_\epsilon\equiv\epsilon\kappa^2 X/2 =\epsilon\kappa^2 C^2/(2r^4A)$. The metric field equations in the Einstein-frame then reads~\cite{AORG:2017}
\begin{equation}
\label{eq:field_eq_EiBI}
\epsilon h^{\mu\alpha}{\cal R}{_{(\alpha\nu)}}(h) = \begin{pmatrix}
\left(1-\dfrac{\lambda+X_\epsilon}{\sqrt{|\Omega|}}\right)I_{3\times 3} & 0 \\
0 & 1-\dfrac{\lambda-X_\epsilon}{\sqrt{|\Omega|}} 
\end{pmatrix} .
\end{equation}

To solve Eq.~\eqref{eq:field_eq_EiBI} one can proceed by using the direct integration of the field equations, just like Wyman did in Ref.~\cite{Wyman:1981}. The key point of this approach is to identify the scalar field as a radial coordinate, which can be done by taking $\phi_x$ as a constant, that is $\phi_x=\upsilon$. With this choice, one can write the following line elements related to the space-time and auxiliary metric, respectively, 
\begin{align}
\label{eq:ds_wyman_EiBI} ds^2 &= -e^{\nu}dt^2+\dfrac{e^\nu}{C_0^2 W^4}dx^2+\dfrac{1}{W^2}(d\theta^2+\sin^2\theta d\varphi^2),\\
\label{eq:ds_wyman_GR} d\tilde{s}^2 &= -e^{\tilde{\nu}}dt^2+\dfrac{e^{\tilde{\nu}}}{C_0^2 \tilde{W}^4}dy^2+\dfrac{1}{\tilde{W}^2}(d\theta^2+\sin^2\theta d\varphi^2),
\end{align}
where $C_0=C/\upsilon$, the metric functions $\nu$ and $W$ depend only on $x$, while $\tilde{\nu}$ and $\tilde{W}$ depend only on $y$. The relation between the metric functions and radial coordinates of the space-time and auxiliary metric are~\cite{AORG:2017}
\begin{align}
&e^{\tilde{\nu}}=\Omega_+e^{\nu},\,\,\,\,\,\,\,\tilde{W}^2=W^2/\Omega_+,\\
&dy=\dfrac{\Omega_-}{\sqrt{|\Omega|}}dx=\dfrac{1}{|\lambda-X_\epsilon|}dx.
\end{align}
From the line elements~\eqref{eq:ds_wyman_EiBI} and~\eqref{eq:ds_wyman_GR}, one obtains that the Klein-Gordon equation~\eqref{eq:kg_equation_canonical} reduces to $\phi_{xx}=0$.
By substituting the line element~\eqref{eq:ds_wyman_GR} in Eq.~\eqref{eq:field_eq_EiBI} and by considering the right-hand side of Eq.~\eqref{eq:field_eq_EiBI} as an effective energy-momentum tensor, $\tau^{\mu}{_\nu}$, one obtains the following set of differential equations
\begin{align}
\label{eq:dfeq1}\tilde{\nu}_{yy} &= -\dfrac{\kappa_0^2\Omega_+^3}{X_\epsilon}\left(1-\dfrac{\lambda+X_\epsilon}{\sqrt{|\Omega|}}\right),\\
\label{eq:dfeq2} &= \tilde{W}_{yy}-\tilde{\nu}_y\tilde{W}_y-\dfrac{\kappa_0^2\Omega_+^3}{2\sqrt{|\Omega|}}\tilde{W},\\
\label{eq:constraintW}\tilde{W}\tilde{W}_{yy}-\tilde{W}^2_y &= -\dfrac{e^{\tilde{\nu}}}{C_0^2}+\dfrac{\kappa_0^2\Omega_+^3}{2X_\epsilon}\left(1-\dfrac{\lambda+X_\epsilon}{\sqrt{|\Omega|}}\right),
\end{align}
where $\kappa_0^2 = \kappa^2\upsilon^2$. There is a technical difficulty in the above set of equations, namely that $X_\epsilon$ depends on $\nu$ and $W$, instead, of $\tilde{\nu}$ and $\tilde{W}$. In order to rewrite $X_\epsilon$ in terms of the tilde functions, one can introduce $\tilde{X}_\epsilon\equiv\epsilon\kappa_0^2C^2\tilde{W}^4e^{-\tilde{\nu}}/2$, such that
\begin{equation}
\label{eq:X_tildeX}
X_{\epsilon}^2(\tilde{X}_\epsilon) = \lambda^2+\dfrac{1}{\sqrt[3]{18}|\tilde{X}_\epsilon|}\left(\tilde{K} - \dfrac{\sqrt[3]{12}}{\tilde{K}}\right),
\end{equation}
with $\tilde{K} = ((12+(9\lambda^2|\tilde{X}_\epsilon|)^2)^{1/2}-9\lambda^2|\tilde{X}_\epsilon|)^{1/3}.$
The negative branch of $\epsilon$ leads to non-singular bouncing solutions in cosmological models~\cite{OORG:2014} and non-singular black holes in electrovacuum scenarios~\cite{ORGSP:2016,Gonzalo:2016}. 

Due to the deep nonlinearity present in the above set of metric field equations, the differential equations~\eqref{eq:dfeq1} and \eqref{eq:dfeq2} can be solved numerically, together with Eq.~\eqref{eq:constraintW}, which is a constraint that $\tilde{\nu}$ and $\tilde{W}$ must satisfy, to obtain the auxiliary metric $h_{\mu\nu}$. Then, by using the deformation matrix, one may recover the space-time metric. A full numerical analysis of this problem was performed in Ref.~\cite{AORG:2017}. The authors found that the areal radius $r=1/W^2(x)$ has a regular minimum, therefore unveiling a throat structure in the space-time. Far from the object, $x\rightarrow 0$, the solution has the same asymptotic behavior as Schwarzschild. The inner region, on the contrary, has a non-asymptotically flat character, where all the algebraic curvature invariants are bounded. The numerical solution for the metric confirms that some approximations are possible in the internal region, that allow to obtain analytical expressions for the metric functions there. Such approximations and the form of the resulting line element are presented in the following section.

\section{Interior solution approximation}\label{sec:appx}
It was noted in Ref.~\cite{MCO:2022} that the asymptotic internal region of~\eqref{eq:ds_quadr}, obtained from the coupling of a QP gravity to a nonlinear scalar field Lagrangian, has the same internal asymptotic structure as the solution discussed in Sec.~\ref{sec:eibi}, which comes from the coupling of EiBI to a canonical scalar field Lagrangian. Far away from the throat in the internal region, the line element of these two space-times can be written as 
\begin{equation}
	\label{eq:approx_metric}
	ds^2 = -\frac{\alpha}{r^4}dt^2 + \frac{\beta}{r^2}dr^2+r^2(d\theta^2 + \sin^2\theta d\varphi^2),
\end{equation}
where $\alpha$ and $\beta$ are constants that depend on the gravitational and matter sectors. The explicit dependence on the parameters of EiBI and QP gravities are, respectively~\cite{AORG:2017,MCO:2022}
\begin{align}
&\alpha_{EiBI}= r_\epsilon^4,&& \beta_{EiBI}=\dfrac{4\sigma^2}{C_0^2},\\
\label{eq:beta_qp}&\alpha_{QP}= \dfrac{2|a|^3\mu^2}{b^2},&& \beta_{QP}=\dfrac{18|a|\mu^2}{(2M-\eta)^2},
\end{align}
where $r_\epsilon \propto \sqrt{|\epsilon|}$ is a constant with dimension of length and $\sigma$ is a combination of some constants that arise during the integration of the field equations.

\subsection{Curvature scalars}
To study some aspects of the asymptotic internal region of the asymmetric wormholes described by the line element~\eqref{eq:approx_metric}, one can compute its curvature invariants. The non-vanishing components of the Riemann tensor, related with the line element~\eqref{eq:approx_metric}, are
\begin{align*}
	R_{t\varphi t\varphi} &= -\frac{2\alpha\sin^{2}\theta}{\beta r^2}, && R_{trtr}=\frac{4\alpha}{r^6},\\
	R_{t\theta t\theta}&=-\frac{2\alpha}{\beta r^2},&& R_{r\theta r\theta} =-1,\\
	R_{r\varphi r\varphi}&= -\sin^2\theta,&& R_{\theta\varphi\theta\varphi}=\frac{r^2(\beta-r^2)\sin^2\theta}{\beta}, \numberthis \label{eq:riemanni}
\end{align*} 
where $R_{\mu \nu \rho \delta}$ denotes the Riemann tensor components of the space-time metric, which should not be confused with the Riemann tensor associated to the independent affine connection, ${\cal R}_{\mu \nu \rho \delta}$. As pointed out in Ref.~\cite{MCO:2022}, the line element~\eqref{eq:approx_metric} has its algebraic curvature invariants finite everywhere in the internal region. For instance the Ricci, Ricci-squared and Kretschmann scalars are given, respectively, by
\begin{align}
	&g^{\mu \nu} R_{\mu \nu}=\frac{2}{r^2}-\frac{6}{\beta},\\
	&R_{\mu \nu} R^{\mu \nu} = \frac{2}{r^4}+\frac{36}{\beta^2},\\
	&R_{\mu \nu \rho \delta} R^{\mu \nu \rho \delta} = \frac{4\left(26{r}^{4}+{\left(\beta-{r}^{2}\right)}^{2}\right)}{{\beta}^{2} {r}^{4}},
\end{align}
which in the asymptotic internal region of the wormhole are constants and bounded by $\lim_{r\to\infty}g^{\mu \nu} R_{\mu \nu}=-6/\beta$, $\lim_{r\to\infty}R_{\mu \nu} R^{\mu \nu}=36/\beta^2$ and $\lim_{r\to\infty} R_{\mu \nu \rho \delta} R^{\mu \nu \rho \delta} = 108/\beta^2$. One can also compute the complete set of algebraic curvature invariants given by Zakhary and McIntosh (ZM), ${\cal K}=(I_1,I_2,I_3,\dots,I_{17})$~\cite{Zakhary:1997}. ZM invariants are a basis for any other algebraic curvature invariant of the Riemann tensor. For example, the Kretschmann scalar in terms of the ZM invariants, in four dimensions, is $R_{\mu \nu \rho \delta} R^{\mu \nu \rho \delta} = I_1+2I_6-I_5^2/3$. It is important to point out that, although the ZM set is complete, it has more than 14 elements, therefore not all of them are independent for all the Petrov and Segre types~\cite{Zakhary:1997}. The relations between the dependent and independent elements are called \textit{syzygies}. We exhibit the non-vanishing ZM invariants of the metric~\eqref{eq:approx_metric} in Appendix~\ref{sec:zm}. In Table~\ref{tab:zm}, we show the asymptotic limit of the ZM invariants, from where one notices that all the 17 ZM invariants are bounded in the asymptotic internal region. Hence, any curvature invariant constructed using the ZM set is also bounded in the asymptotic internal region.

\begin{table}[!h]
\centering \caption{Asymptotic limit of the ZM invariants of the line element~\eqref{eq:approx_metric} in the internal region.}\label{tab:zm}
\vskip 10pt
\begin{tabular}{@{}ccccccc@{}}
\hline \hline
\hspace{0.5cm}ZM invariant\hspace{0.5cm}&\hspace{0.5cm} Degree \hspace{0.5cm} &\hspace{0.5cm} Asymptotic limit ($r\to\infty$) \hspace{0.5cm}\\
 $I_1$               &2          & $48/\beta ^2$         \\
$I_2$               &2          & $0$         \\
$I_3$               &3          & $-96/\beta ^3$         \\
$I_4$               &3          & $0$         \\
 $I_5$               &1          & $-6/\beta $         \\
$I_6$               &2          & $36/\beta ^2$         \\
$I_7$               &3          & $-216/\beta ^3$         \\
$I_8$               &4          & $1296/\beta ^4$         \\
 $I_9$               &3          & $0$         \\
$I_{10}$               &3          & $0$         \\
$I_{11}$               &4          & $216/\beta ^4$         \\
$I_{12}$               &4          & $0$         \\
 $I_{13}$               &5          & $0$         \\
$I_{14}$               &5          & $0$         \\
$I_{15}$               &4          & $27/2 \beta ^4$         \\
$I_{16}$               &5          & $-27/2 \beta ^5$         \\
$I_{17}$               &5          & $0$         \\
\hline \hline
\end{tabular}
\label{table1}
\end{table}

\subsection{Geodesic analysis}
To further investigate the geometry of the internal region of the asymmetric wormholes generated by scalar fields in RBGs with asymptotic line element~\eqref{eq:approx_metric}, it is useful to study the geodesics on it. The motion of free point-like particles are encoded in the Lagrangian $2{\cal L} = \dot{s}^2 = k$, where the overdot denotes a derivative with respect to an affine time $\tau$ and $k=-1$ denotes a massive particle, while $k=0$ corresponds to light-rays. From the symmetries of the line element~\eqref{eq:approx_metric}, the Lagrangian has a time translation symmetry and also a rotational symmetry. Therefore, there are two conserved quantities along the geodesics: the energy of the particle, $E$, due to $\partial {\cal L}/\partial t =0$, and the angular momentum of the particle, $L$, due to $\partial {\cal L}/\partial \varphi =0$. (For light-rays it is convenient to define the impact parameter via the ratio $L/E$.) By using the Euler-Lagrange equation, one obtains that geodesics in the equatorial plane $(\theta=\frac{\pi}{2})$, in the asymptotic internal region of the asymmetric wormhole, satisfy
\begin{equation}
\label{eq:geo_eq}
	\dfrac{\alpha \beta}{r^6}\dot{r}^2 = E^{2}-\dfrac{\alpha}{r^4}\left(\dfrac{L^2}{r^2}-k\right).
\end{equation}
For massless particles, radial geodesics (with $L=0$) in the internal region become
\begin{equation}
\label{eq:radial_geo_k0}
\dfrac{\alpha\beta}{4}\dot{u}^2 = E^2,
\end{equation}
while for massive particles we have
\begin{equation}
\label{eq:radial_geo_km1}
\dfrac{\alpha\beta}{4}\dot{u}^2 = E^2-\alpha u^2,
\end{equation}
where we defined $u(\tau)\equiv 1/r(\tau)^2$.

By solving Eq.~\eqref{eq:radial_geo_k0} one finds that radial light-rays follow 
\begin{equation}
\label{eq:light_ray_rad_geo}
u(\tau) = \pm\dfrac{2E\tau}{\sqrt{\alpha\beta}}+u_0,
\end{equation}
where the $\pm$ sign denotes ingoing ($-$) and outgoing ($+$) geodesics, with $u_0=1/r_0^2$ determined by the initial position of the massless particle.
We notice that for radial geodesics propagating towards the asymptotic internal region (the minus-sign solution of Eq.~\eqref{eq:light_ray_rad_geo}), $u(\tau)$ vanishes at a finite value of $\tau$, namely,
\begin{equation}
\tau_f = \dfrac{\sqrt{\alpha\beta}}{2E}u_0 \ .
\end{equation}
This means that $r(\tau)$ goes to infinity after a finite affine time, implying that radial null geodesics are incomplete.

The radial motion of massive particles is found by solving Eq.~\eqref{eq:radial_geo_km1}, whose solutions can be easily obtained by just taking a time derivative of that equation, finding that
\begin{equation}
\label{eq:massive_particle_rad_geo}
u(\tau) = \dfrac{E}{\sqrt{\alpha}}\cos\left(\frac{2(\tau-\tau_0)}{\sqrt{\beta}}\right),
\end{equation}
where $\tau_0$ denotes the instant at which $r_0^2=\sqrt{\alpha}/E$. The asymptotic region $r\to \infty$ is obtained at time $\tau=\tau_0+(\pi\sqrt{\beta}/4)$, beyond which the solution becomes negative and no longer represents a physical situation. Thus, radial time-like geodesics are also incomplete. From now on, we chose the integration constants $u_0$ and $\tau_0$ such that the singularity is reached at $\tau=0$ (therefore, the affine parameter is negative). 

It is also worth pointing out that, regardless of $k$ or $L$, Eq.~\eqref{eq:geo_eq} for large $r$ can be approximated by~\cite{MCO:2022}
\begin{equation}
\label{eq:geo_app}
\dot{r}^2=\dfrac{E^2}{\alpha\beta}r^6+{\cal O}(r^2),
\end{equation}
which indicates that all geodesics degenerate into incomplete radial null geodesics (see Eq.~\eqref{eq:radial_geo_k0}). Therefore, asymmetric wormholes with an internal region described by the line element~\eqref{eq:approx_metric} are (null- and time-like) geodesically incomplete. Accordingly, they represent singular space-times despite having bounded algebraic curvature invariants everywhere.

\section{Parallelly propagated curvature singularities}\label{sec:pp}
Geodesically incomplete space-times can be classified as having  curvature or non-curvature singularities~\cite{Ellis:1979}. If a space-time has a scalar polynomial of the Riemann tensor unbounded on an incomplete curve, one says that the space-time has a scalar (polynomial) curvature singularity. The Wyman solution of Eq.~(\ref{eq:wyman}) is an example of this sort of singularity, since its curvature invariants blow up along incomplete geodesics. If any component of the curvature tensor in a basis parallelly propagated along an incomplete curve is unbounded, one says that the space-time has a parallelly propagated curvature singularity. (For example, if the space-time presents unbounded tidal forces.) Other types of singularities without curvature nature can also be found, such as the quasiregular singularities~\cite{Clarke:1977,Vickers:1987,Konkowski:1985}.


A remarkable feature of the class of wormholes studied in this paper is that all their algebraic curvature invariants are finite, despite the presence of incomplete causal geodesics. This suggests that the singularities on these wormholes are of a non-scalar nature. In order to probe the nature of these singularities, let us investigate how two nearby geodesics behave in these wormholes, specifically by studying the tidal forces exerted by them. 

\subsection{Tidal forces in the internal region of non-regular wormholes}

In a curved space-time, two nearby geodesics may move toward or away from each other. Mathematically, the relative acceleration between two nearby
particles is described by the so-called geodesic deviation equation,
namely~\cite{Pirani_rep, dinverno}
\begin{equation}
	\label{eq:gdeviation}
	\frac{D^{2}\xi^{\mu}}{D\tau^{2}}={R^{\mu}}_{\rho\nu\beta}T^{\rho}T^{\nu}\xi^{\beta},
\end{equation}
where $\xi^{\mu}$ represents the separation vector between (infinitesimally) nearby geodesics (these vectors are also known as Jacobi fields, and in Sec.~\ref{sec:str} we make use of this nomenclature) and $T^{\mu}$ are the components of the tangent vector to the geodesics. As we saw, both time-like and null-like radial geodesics are incomplete in the interior region of the asymmetric wormhole described by Eq.~\eqref{eq:approx_metric}. Therefore, let us investigate how two nearby time-like geodesics propagating radially toward the asymptotic region behave according to Eq.~\eqref{eq:gdeviation}. For this aim, it is convenient to introduce a \textit{vierbein}, that is, a set of orthonormal vectors, that follows the radial time-like geodesics. One of the vectors in the vierbein is chosen to represent the tangent vector to the geodesics~\cite{Crispino:2016,Goel:2015}, which for time-like geodesics is given by $\boldsymbol e_{\hat{0}} = \dot{t}\, \partial_t+\dot{r}\,\partial_r $. Here the hat index labels each vector in the vierbein, $\partial_t$ and $\partial_r$ together with $\partial_\theta$ and $\partial_\varphi$ are the basis vectors associated to the coordinate system, $\dot{t}=Er^4/\alpha$ and, from Eq.~\eqref{eq:geo_eq}, the radial component of the four-velocity is
\begin{equation}
\label{eq:radial_vel}
\dot{r}=\pm r^{3}\sqrt{\dfrac{E^2 - \dfrac{\alpha}{r^4}}{\alpha\beta}},
\end{equation}
with the positive and negative signs being related to ingoing and outgoing geodesics, respectively. The other three vectors can be found by applying the orthonormality condition, $g_{\mu\nu}e_{\hat{a}}{^{\mu}}e_{\hat{b}}{^{\nu}}=\eta_{\hat{a}\hat{b}}$, where $\eta_{\hat{a}\hat{b}}=(-1,1,1,1)$ are the components of the Minkowski metric in Cartesian coordinates. Therefore, the vectors of the vierbein are
\begin{align}
	\boldsymbol e_{\hat{0}}&= \frac{Er^{4}}{\alpha}\,\partial_t \pm r^{3}\sqrt{\frac{E^2 - \dfrac{\alpha}{r^4}}{\alpha\beta}}\,\partial_r,\\
	\boldsymbol e_{\hat{1}}&= \pm\frac{r^4}{\alpha}\sqrt{E^2 - \dfrac{\alpha}{r^4}}\,\partial_t + \frac{Er^3}{\sqrt{\alpha\beta}}\,\partial_r ,\\
	\boldsymbol e_{\hat{2}}&= \dfrac{1}{r}\partial_{\theta},\\
	\boldsymbol e_{\hat{3}}&= \dfrac{1}{r\sin\theta}\partial_\varphi.
\end{align}

In the orthonormal frame, the geodesic deviation equation~\eqref{eq:gdeviation} can be written as~\cite{Crispino:2016}
\begin{equation}
	\frac{D^{2}\xi^{\hat{a}}}{D\tau^{2}}={K^{\hat{a}}}_{\hat{b}}\xi^{\hat{b}}, 
\end{equation}
where ${K^{\hat{a}}}_{\hat{b}}\equiv{R^{\hat{a}}}_{\hat{0}\hat{0}\hat{b}}={R^{\mu}}_{\rho\nu\beta}e^{\hat{a}}_{\mu}e^{\rho}_{\hat{0}}e^{\nu}_{\hat{0}}e^{\beta}_{\hat{b}}$ are the components of the tidal tensor in the parallelly propagated frame $\{\boldsymbol e_{\hat{a}}\}$. From the computed components of the Riemann tensor~\eqref{eq:riemanni}, one finds that the non-vanishing tidal tensor components are given by
\begin{align}
	\label{eq:rad_tid}{K^{\hat{1}}}_{\hat{1}}&=-\frac{4}{\beta},\\
	\label{eq:ang_tid}{K^{\hat{2}}}_{\hat{2}}&={K^{\hat{3}}}_{\hat{3}}=\frac{3r^4E^2-\alpha}{\alpha\beta}.	
\end{align}
By substituting the above equations in Eq.~\eqref{eq:gdeviation}, one obtains the expressions for the relative acceleration between two nearby geodesics, that is, the radial and angular tidal forces in the asymptotic internal region of the wormhole. 
One notices that, since $\beta$ is positive in both the EiBI and the QP models \cite{AORG:2017,MCO:2022}, the radial tidal force, ${K^{\hat{1}}}_{\hat{1}}$ has a negative constant value in the asymptotic internal region of the asymmetric wormhole. On the other hand, the angular tidal forces, ${K^{\hat{2}}}_{\hat{2}}$ and ${K^{\hat{3}}}_{\hat{3}}$, are unbounded in the internal region of the wormhole, presenting a positive term proportional to $r^4$.  

As previously mentioned, if any component of the Riemann tensor in a vierbein associated to an incomplete curve is unbounded, the space-time singularity is considered a parallelly propagated curvature singularity. In the case of the asymmetric wormholes under study, the components ${R^{\hat{2}}}_{\hat{0}\hat{0}\hat{2}}$ and ${R^{\hat{3}}}_{\hat{0}\hat{0}\hat{3}}$ become unbounded along incomplete radial time-like geodesics. Accordingly, while an observer in the internal region of the wormhole might measure finite curvature scalars, they would experience infinite angular stretching. 

For the sake of completeness, we also present the tidal tensor components for a static observer whose vierbein is given by
\begin{align}
\boldsymbol \lambda_{\hat{0}}&= \frac{r^2}{\sqrt{\alpha}}\partial_t, \qquad
\boldsymbol \lambda_{\hat{1}}=\frac{r}{\sqrt{\beta}}\partial_r,\\
\boldsymbol \lambda_{\hat{2}}&= \dfrac{1}{r}\partial_{\theta}, \qquad \ \ \ \boldsymbol \lambda_{\hat{3}}= \dfrac{1}{r\sin\theta}\partial_\varphi.
\end{align}
The non-vanishing tidal tensor components for the static frame are:
\begin{align}
	\label{eq:rad_tid_static}{\widetilde{K}^{\hat{1}}}_{\ \ \hat{1}}&=-\frac{4}{\beta},\\
	\label{eq:ang_tid_static}{\widetilde{K}^{\hat{2}}}_{\ \ \hat{2}}&=\frac{2}{\beta}.	
\end{align}
We notice that the tidal forces, as measured by a static observer, are bounded regardless of the value of $r$. The observation of finite tidal forces by a static observer is consistent with the presence of a parallelly propagated singularity, since a static observer in the asymptotic internal region is subjected to an acceleration, hence it is not parallelly propagated. Curiously, the radial tidal force is the same in the parallelly propagated and static frames.

In the following section we investigate how Jacobi fields and volume elements evolve in the internal region when approaching the singularity, in order to characterize the \textit{strength} of the parallelly propagated curvature singularity in this wormhole scenario.

\subsection{Strength of the singularity}\label{sec:str}

In order to classify the various types of singularities, criteria about their strength and impact on observers have been established in the literature. The first and more intuitive criterion is the characterization of a \textit{strong} curvature singularity as that case in which all objects moving towards it are compressed to zero volume. This was mathematically formalized by Tipler~\cite{Tipler:1977a,Tipler:1977b} and further refined by Clarke and Kr\'olak~\cite{Clarke:1985}. One can establish that, according to Tipler's definition, a necessary condition for volume elements following incomplete time-like geodesics to be crushed into zero volume at a point (with affine parametrization $\tau=0$) is that for some component ${R^{\hat{a}}}_{\hat{0}\hat{b}\hat{0}}$ of the Riemann tensor in the vierbein basis, the quantity
\begin{equation}
\label{eq:tipler}
{I^{\hat{a}}}_{\hat{b}}(\tau) = \int_{\tau_1}^{\tau}d\tau'\int_{\tau_1}^{\tau'}d\tau'' |{R^{\hat{a}}}_{\hat{0}\hat{b}\hat{0}}|
\end{equation}
does not converge as $\tau\to 0^{-}$, with $\tau_1<0$. On the other hand, the necessary condition for strong curvature singularities according to Kr\'olak's definition is that
\begin{equation}
\label{eq:krolak}
{J^{\hat{a}}}_{\hat{b}}(\tau) = \int_{\tau_1}^{\tau}d\tau' |{R^{\hat{a}}}_{\hat{0}\hat{b}\hat{0}}|
\end{equation}
does not converge as $\tau\to 0^{-}$. Sufficient conditions for both definitions are obtained by replacing $|{R^{\hat{a}}}_{\hat{0}\hat{b}\hat{0}}|$ by $R_{\mu\nu}T^{\mu}T^{\nu}$ in Eqs.~\eqref{eq:tipler} and~\eqref{eq:krolak}, together with the time-like or null convergence condition~\cite{Clarke:1985}, $R_{\mu\nu}T^{\mu}T^{\nu}\geq 0$, where $R_{\mu\nu}$ are the components of the Ricci tensor of the space-time metric. In order to make clearer the physical meaning of a strong singularity, other characterizations were introduced, in particular, the notion of \textit{deformationally} strong singularity was proposed in Refs.~\cite{nolan,ori}, where it also includes the case in which any of the Jacobi fields is unbounded. This can lead, for instance, to a divergent volume element or a finite volume element at the singularity, which according to Tipler's definition would be a \textit{weak} curvature singularity. It was shown in Ref.~\cite{ori} that the necessary condition for Tipler's strong singularities is also a necessary condition for deformationally strong singularities. Therefore, for strong singularities, the integral~\eqref{eq:tipler} does not converge as $\tau\to 0^{-}$. One should bear in mind that any Tipler's strong singularity is also a deformationally strong singularity, though the opposite is not true.\\

Let us now check if the necessary condition for strong singularities is fulfilled by the singularity in the internal region of the wormhole. By substituting, for instance, ${R^{\hat{2}}}_{\hat{0}\hat{2}\hat{0}}=-{K^{\hat{2}}}_{\hat{2}}$ in Eq.~\eqref{eq:tipler} and using that near the singularity ($r\to\infty$), $r\approx (\alpha\beta)^{1/4}/\sqrt{2E|\tau|}$, one obtains
\begin{equation}
\label{eq:cond}
{I^{\hat{a}}}_{\hat{b}}(\tau) = \int_{\tau_1}^{\tau}d\tau'\int_{\tau_1}^{\tau'}d\tau'' \left(\frac{3}{4\tau^2}-\frac{1}{\beta}\right),
\end{equation}
where we are also taking into account the fact that ${R^{\hat{2}}}_{\hat{0}\hat{2}\hat{0}}<0$ near the singularity. Actually, this is true for any $-\sqrt{3\beta}/2<\tau<0$. One can verify that Eq.~\eqref{eq:cond} does not converge as $\tau\to 0^{-}$, since the integral diverges with $-3\ln|\tau|/4$ near the singularity. Therefore, the necessary condition is satisfied.

 By analyzing the component $R_{\hat{0}\hat{0}}$ of the Ricci tensor in the vierbein, which near the singularity is
\begin{equation}
R_{\hat{0}\hat{0}} \approx \dfrac{6}{\beta}-\frac{3}{2\tau^2},
\end{equation}
one notices that the time-like convergence condition, $R_{\hat{0}\hat{0}}\geq 0$, is not fulfilled when approaching the singularity, which means that the sufficient condition for the singularity to be a Tipler's strong singularity does not hold. Therefore, we have to take another approach to identify this singularity.

In order to characterize the strength of the singularity in the asymptotic internal region of the asymmetric wormhole, we have to study the solutions of the geodesic deviation equation. We are interested in investigating if the Jacobi fields are bounded when they approach the singularity, and how volume elements behave in its neighborhood. As already mentioned, a Jacobi field $\boldsymbol \xi$ is any smooth vector field that satisfies Eq.~\eqref{eq:gdeviation} along a causal geodesic. Since Eq.~\eqref{eq:gdeviation} is a second order differential equation, there are six independent Jacobi fields, depending on the values of $\xi^{\hat{a}}$ and $\dot{\xi}^{\hat{a}}$ at some point $\tau_i$. (In this section we are using the overdot to denote $D/D\tau$.) The components of any Jacobi field may be expressed in terms of the initial values $\xi^{\hat{a}}(\tau_i)$ via $\xi^{\hat{a}}(\tau)={{\cal A}^{\hat{a}}}_{\hat{b}}(\tau)\xi^{\hat{b}}(\tau_i)$ if all $\xi^{\hat{a}}(\tau_i)\neq 0$, or via $\xi^{\hat{a}}(\tau)={{\cal B}^{\hat{a}}}_{\hat{b}}(\tau)\dot{\xi}^{\hat{b}}$ if all $\xi^{\hat{a}}(\tau_i)=0$. The components of the $3\times 3$ matrices ${{\cal A}^{\hat{a}}}_{\hat{b}}(\tau)$ and ${{\cal B}^{\hat{a}}}_{\hat{b}}(\tau)$ go to ${\delta^{\hat{a}}}_{\hat{b}}$ and zero, respectively, at $\tau_i$. With three independent Jacobi fields $\boldsymbol \xi_{(i)}={\xi_{(i)}}^{\hat{a}}\boldsymbol e_{\hat{a}}$, with $i=1,2,3$, one can write the volume 3-form along a given geodesic as~\cite{nolan}
\begin{equation}
\label{eq:volume_form}
V(\tau) = {\boldsymbol \xi}_{(1)}\wedge{\boldsymbol \xi}_{(2)}\wedge{\boldsymbol \xi}_{(3)}.
\end{equation}
One can relate this volume element with the determinant of the matrices ${{\cal A}^{\hat{a}}}_{\hat{b}}(\tau)$ and ${{\cal B}^{\hat{a}}}_{\hat{b}}(\tau)$ and some initial data of the volume at $\tau_i$. In particular, when all $\xi^{\hat{a}}(\tau_i)=0$, the volume may be written as  $V(\tau) = \det|{{\cal B}^{\hat{a}}}_{\hat{b}}(\tau)|.$

Let us follow Ref.~\cite{nolan} to compute the functional dependence of the volume form~\eqref{eq:volume_form}. For this aim, it is convenient to rewrite the line element~\eqref{eq:approx_metric} using null coordinates, that is,
\begin{equation}
ds^2=-\dfrac{\beta}{(v-u)^2}dvdu+r^2(u,v)(d\theta^2 + \sin^2\theta d\varphi^2),
\end{equation}
where we defined the null coordinates $v\equiv t+r_\star$ and $u\equiv t-r_\star$ in terms of the radial coordinate $r_\star= r^2\sqrt{\beta/4\alpha}$. 
In the above coordinate system, namely $(u,v,\theta,\varphi)$, one may write the tangent vector of an arbitrary time-like curve $\gamma(\tau)$ as $\boldsymbol T=h\partial_u+(\beta h)^{-1}(v-u)^2\partial_v$, such that $T^\mu T_\mu = -1$, where $h\equiv h(u,v)$ is an arbitrary function of the null coordinates $u$ and $v$. If $\gamma(\tau)$ is a geodesic, its tangent vector must satisfy $T^\mu \nabla_\mu T^\nu = 0$, therefore one finds that $h(u,v)$ satisfies $\partial_\nu h+\beta h^2(\partial_uh-2h/(u-v))/(u-v)^2=0$. 

Let us take three orthogonal vectors to the unit tangent vector of a time-like curve, $\boldsymbol T$, namely
\begin{align}
\boldsymbol \xi_{(1)} &= a(u,v)\left[h(u,v)\partial_u-\dfrac{(v-u)^2}{\beta h(u,v)}\partial_v\right],\\
\boldsymbol \xi_{(2)} &= p(u,v)\partial_\theta,\\
\boldsymbol \xi_{(3)} &= q(u,v)(\sin\theta)^{-1}\partial_\varphi,
\end{align}
as Jacobi fields candidates.
The functions $a$, $p$ and $q$ are found by requiring that those fields satisfy the geodesic deviation equation. In terms of the affine parameter $\tau$, the function $p$ (and equivalently $q$) must satisfy the equation
\begin{equation}
r\ddot{p}+2\dot{r}\dot{p}=0,
\end{equation}
whose solution can be cast as
\begin{equation}
p(\tau)=p_1+p_0\int\dfrac{d\tau}{r^2(\tau)},
\end{equation}
where $p_0$ and $p_1$ are integration constants. The function $a$ must satisfy the equation
\begin{equation}
\label{eq:a}
\ddot{a}+\dfrac{4}{\beta}a=0,
\end{equation}
that is, $a$ satisfies a differential equation with the same coefficients as the radial tidal force equation in the vierbein (see Eq.~\eqref{eq:rad_tid}). The general solution to Eq.~\eqref{eq:a} is
\begin{equation}
a(\tau) = a_1\cos\left(2\tau/\sqrt{\beta}\right)+a_2\sin\left(2\tau/\sqrt{\beta}\right),
\end{equation}
where $a_1$ and $a_2$ are integration constants.

One can evaluate the norm of the Jacobi fields, finding that $|\boldsymbol \xi_{(1)}|=|a|$, $|\boldsymbol \xi_{(2)}|=r|p|$ and $|\boldsymbol \xi_{(3)}|=r|q|$. Therefore, one finds that the norm of the volume 3-form is simply given by
\begin{equation}
\label{eq:vol_norm}
|V(\tau)| = |apq|r^2.
\end{equation}
In order to analyze how the volume evolves along the geodesic, one may consider two sorts of initial conditions, namely: (IC-I) $\boldsymbol \xi_{(i)}(\tau_1)=0$ and $\dot{\boldsymbol \xi}_{(i)}(\tau_1)>0$ or (IC-II) $\boldsymbol \xi_{(i)}(\tau_1)>0$ and $\dot{\boldsymbol \xi}_{(i)}(\tau_1)=0$, with $\tau_1<0$. Intuitively, the initial condition IC-I corresponds to a body (made by dust) \textit{exploding} from a point $r_i=r(\tau_i)$. The initial condition IC-II, on the other hand, corresponds to releasing a body initially at rest.

Let us consider the initial condition IC-I (one can check that by considering IC-II, the analysis will lead to the same qualitative results), namely
\begin{align}
a(\tau_1)&=0,\,\,\, \dot{a}(\tau_1)=C_1,\\
p(\tau_1)&=0,\,\,\, \dot{p}(\tau_1)=C_2,\\
q(\tau_1)&=0,\,\,\, \dot{q}(\tau_1)=C_3,
\end{align}
with $C_i>0$ $(i=1,2,3)$. 
In the asymptotic internal region, $r\to\infty$, $r\approx (\alpha\beta)^{1/4}/\sqrt{2E|\tau|}$. Thus, the behaviors of the functions $a$, $p$ and $q$ near the singularity ($\tau\to 0^{-}$) are, respectively,
\begin{align}
a(\tau)&\approx \dfrac{\tilde{C}_0}{2}+\tilde{C}_1|\tau|+\dfrac{\tilde{C}_0}{\sqrt{\beta}}\tau^2,\\
p(\tau)&\approx\dfrac{C_2}{|\tau_1|}(\tau_1^2-\tau^2),\\
q(\tau)&\approx\dfrac{C_3}{|\tau_1|}(\tau_1^2-\tau^2),
\end{align}
where $\tilde{C}_0$ and $\tilde{C}_1$ are constants that depend on $C_1$, $|\tau_1|$ and $\beta$. One can see that as the singularity is approached, $a$, $p$ and $q$ go to constant values, namely $a_0=\tilde{C}_0/2$, $p_0=C_2\tau_1^2/|\tau_1|$ and $q_0=C_3\tau_1^2/|\tau_1|$, respectively. From this, one sees that $\boldsymbol \xi_{(2)}$ and $\boldsymbol \xi_{(3)}$ are unbounded, because their norm, respectively $|p|r$ and $|q|r$, diverges near the singularity. This characterizes the singularity as a deformationally strong singularity~\cite{ori}. The behavior of the volume~\eqref{eq:vol_norm} near the singularity is
\begin{equation}
\label{eq:vol_near_sing}
|V(\tau)| \approx \dfrac{|a_0 p_0 q_0|\sqrt{\alpha\beta}}{2E|\tau|}.
\end{equation}
As the singularity is approached, the volume behaves as $|V(\tau)|\sim |\tau|^{-1}$ and diverges as $\tau\to 0^-$. This divergence is caused by the infinite positive tidal force experienced in the angular directions, while the radial tidal force remains constant at large distances within the internal region. Thus, observers following radial geodesics towards the asymptotic internal region are ripped apart and their story ends in a deformationally strong singularity.

\section{Tidal forces in asymmetric wormholes of quadratic Palatini gravity}\label{sec:tidal_QP}
In the previous section we investigated the tidal forces in the internal region of the asymmetric wormholes supported by scalar fields in EiBI and QP gravities. In this section, for completeness, we present a full tidal force description of the QP gravity wormhole. We have chosen QP instead of EiBI gravity wormhole due to the fact that, despite being cumbersome, in the former we have an analytical expression for the space-time metric everywhere.

\begin{figure}[!h]
\center
	\includegraphics[width=\columnwidth]{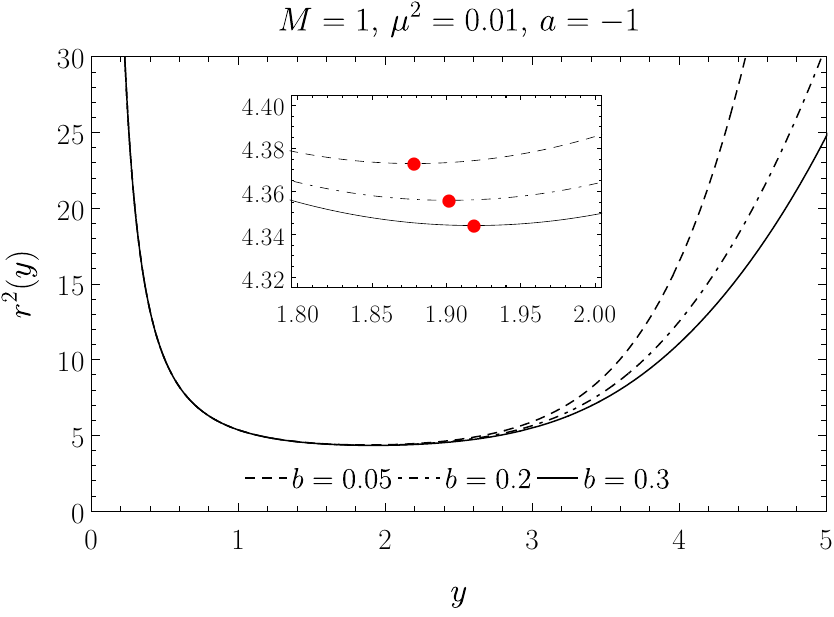}
	\caption{Behavior of the areal radius squared, $r^2(y)$, of the QP asymmetric wormhole for some values of the theory parameter $b$. The red dots in the figure represent the location of the wormhole throat, which is identified by the local minimum of the areal radius.}
	\label{fig:r2M1}
\end{figure}

\begin{figure}[!h]
\center
	\includegraphics[width=\columnwidth]{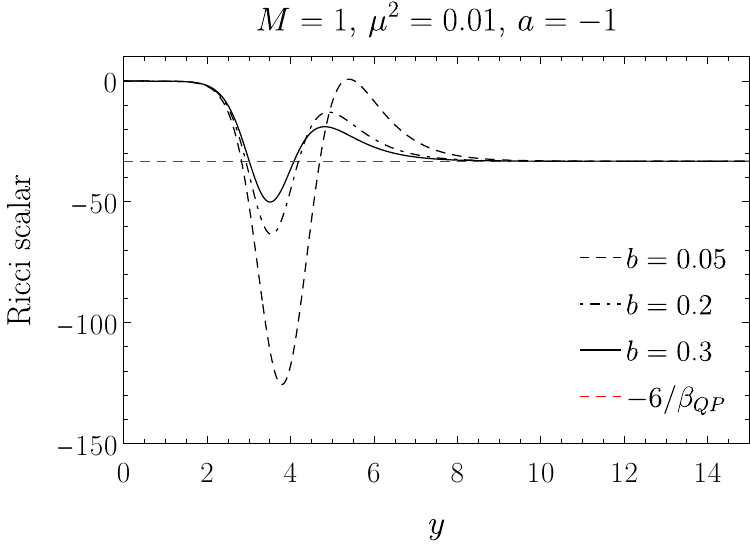}\\\includegraphics[width=\columnwidth]{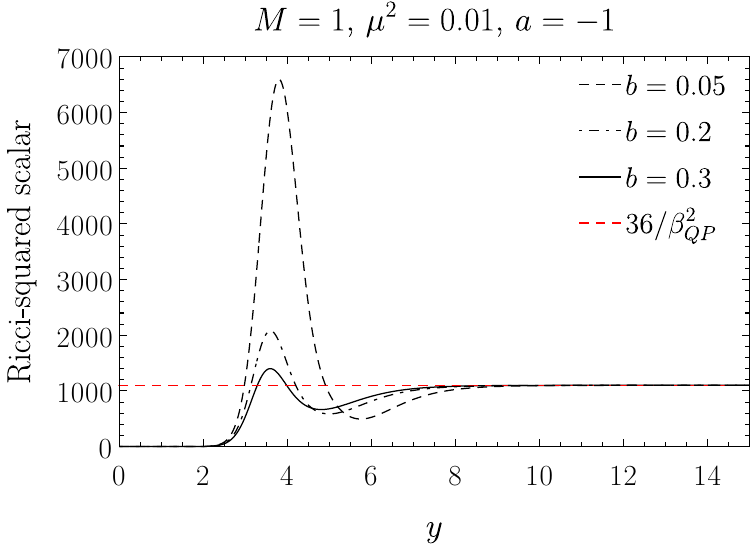}\\\includegraphics[width=\columnwidth]{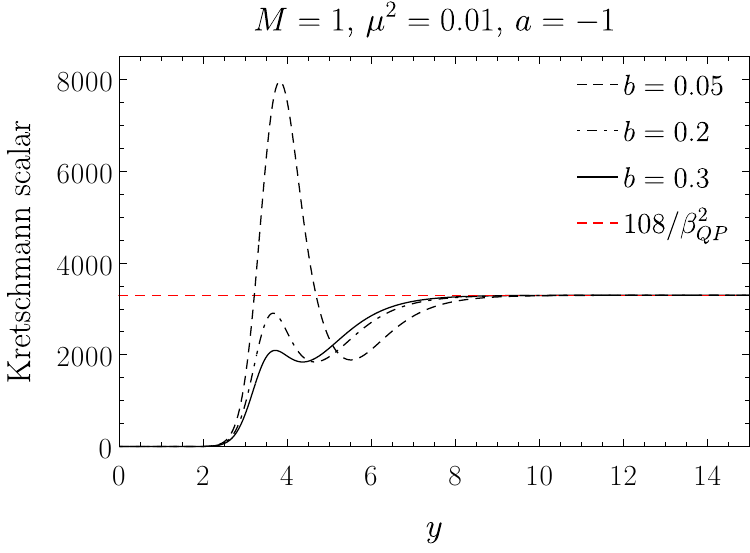}
	\caption{Behavior of the Ricci (top), Ricci-squared (middle) and Kretschmann (bottom) scalars of the QP asymmetric wormhole for some choices of the theory parameter $b$. In the outer region far from the wormhole's throat, the scalars all vanish, indicating that the space-time is asymptotically flat. In the asymptotic internal region these scalars go to finite values, hence they are bounded everywhere. The red dashed lines are the asymptotic values found using the approximated line element~\eqref{eq:approx_metric}.}
	\label{fig:scalars}
\end{figure}

Let us write the line element~\eqref{eq:ds_quadr} as
\begin{equation}
\label{eq:ds_asymwh}
ds^2=-A(y)dt^2+B(y)dy^2+r^2(y)(d\theta^2 + \sin^2\theta d\varphi^2),
\end{equation}
where
\begin{align}
A(y)&=\dfrac{e^{\nu}}{\sqrt{f_{\cal R}F_X}},\\
B(y)&=\dfrac{e^{\nu}W^{-4}}{\sqrt{f_{\cal R}F_X}}+\dfrac{2\mu^2b}{F_X},\\
r^2(y)&=\dfrac{W^{-2}}{\sqrt{f_{\cal R}F_X}}, 
\end{align}
where the definitions of $\nu$, $W$, $f_{\cal R}$ and $F_X$ where provided in Sec.~\ref{sec:f(R,Q)}. In Fig.~\ref{fig:r2M1} we show the behavior of the areal radius squared for this space-time. We are fixing the theory parameter $a=-1$, the mass $M=1$ and the amplitude of the field $\zeta=0.1$ (adopting $\kappa^2=1$, it corresponds to $\mu^2=0.01$), and considering $b=0.05$, $0.2$ and $0.3$. The minimum in this function identifies the throat of the wormhole~\cite{MCO:2022} that connects the two asymptotic regions of the space-time. For $b=0.05$, $0.2$ and $0.3$, the throats are approximately located at $y_{th}\approx 1.877$, $1.901$ and $1.918$, respectively. The limit $y\to 0$ represents the original asymptotically flat region, which rapidly converges to the Wyman solution far from the naked singularity. On the other hand, the limit $y\to\infty$ corresponds to the internal non-asymptotically flat space. In Fig.~\ref{fig:scalars}, we plot some of the curvature invariants of the asymmetric wormhole, from where one notices that the Ricci, Ricci-squared, and Kretschmann scalars, respectively $g^{\mu\nu}R_{\mu\nu}$, $R^{\mu\nu}R_{\mu\nu}$ and $R^{\mu\nu\rho\delta}R_{\mu\nu\rho\delta}$, are finite everywhere. We also notice that the asymptotic values of the curvature invariants in the inner region of the wormhole are in very good agreement with the ones found using the approximation discussed in Sec.~\ref{sec:appx}, even for not too large values of $y$.

In order to compute the tidal forces in the whole space-time, we introduce a vierbein attached to a radial time-like geodesic, namely~\cite{Crispino:2016,Goel:2015},
\begin{align}
\boldsymbol e_{\hat{0}}&= \frac{E}{A(y)}\,\partial_t \pm \sqrt{\frac{E^2 - A(y)}{A(y)B(y)}}\,\partial_y,\\
\boldsymbol e_{\hat{1}}&= \pm \frac{1}{A(y)}\sqrt{E^2 - A(y)}\,\partial_t + \frac{E}{\sqrt{A(y)B(y)}}\,\partial_y ,\\
\boldsymbol e_{\hat{2}}&= \dfrac{1}{r(y)}\partial_{\theta},\\
\boldsymbol e_{\hat{3}}&= \dfrac{1}{r(y)\sin\theta}\partial_\varphi,
\end{align}
where the plus and minus signs refer to ingoing or outgoing geodesics, respectively. Again, we are identifying the zeroth vector of the basis as the unit tangent vector to the geodesic (four-velocity), and using the orthonormality condition, $g_{\mu\nu}e_{\hat{a}}{^{\mu}}e_{\hat{b}}{^{\nu}}=\eta_{\hat{a}\hat{b}}$, to determine the other vectors.

By computing the independent components of the Riemann tensor associated with the line element~\eqref{eq:ds_asymwh}, one can find the components of the tidal tensor via ${K^{\hat{a}}}_{\hat{b}} = {R^{\mu}}_{\nu\alpha\beta}e_{\mu}^{\hat{a}}e^{\nu}_{\hat{0}}e^{\alpha}_{\hat{0}}e^{\beta}_{\hat{b}}$, namely,
\begin{align*}
{K^{\hat{1}}}_{\hat{1}} &= \frac{A(y) A'(y) B'(y)+B(y) \left(A'(y)^2-2 A(y) A''(y)\right)}{4 A(y)^2 B(y)^2}\numberthis,\\
{K^{\hat{2}}}_{\hat{2}}&={K^{\hat{3}}}_{\hat{3}} = \frac{2 A(y) B(y) \left(E^2-A(y)\right) r''(y)}{2 A(y)^2
   B(y)^2 r(y)}+\\
   &-\dfrac{r'(y) \left(E^2 B(y) A'(y)+A(y) \left(E^2-A(y)\right) B'(y)\right)}{2 A(y)^2
   B(y)^2 r(y)},\numberthis
\end{align*}
where the primes denote derivatives with respect to $y$. Far from the throat, in the outside region ($y\to 0$), $W(y)\approx y$, such that the kinetic term $\kappa^2 Z \approx \mu^2 y^4\to 0$. In the limit $Z\to 0$, one recovers $X\approx Z$~\cite{MCO:2022}, such that $f_{\cal R}\approx 1$ and $f_X\approx 1$, hence the 
line element~\eqref{eq:ds_asymwh} reduces to the Wyman solution. The tidal forces for the Wyman space-time are
\begin{align*}
{K^{\hat{1}}}_{\hat{1}} &= -\frac{2 M e^{-2 M y} \sinh ^4(\eta  y) (M-\eta  \coth (\eta 
   y))}{\eta ^4}\numberthis,\\
{K^{\hat{2}}}_{\hat{2}}&={K^{\hat{3}}}_{\hat{3}}=-\frac{M e^{-2 M y} \sinh ^3(\eta  y) \cosh (\eta  y)}{\eta ^3}\\&\frac{e^{-2 M y} \sinh ^4(\eta  y) \left(E^2 (M-\eta ) (\eta +M) e^{2
   M y}+\eta ^2\right)}{\eta ^4}.\numberthis
\end{align*}	
Thus, one can check that $y\to 0$ leads to zero tidal forces in the radial and angular directions, as expected, given that $y\to 0$ corresponds to an asymptotically flat space. 

\begin{figure}[!h]
\center
	\includegraphics[width=\columnwidth]{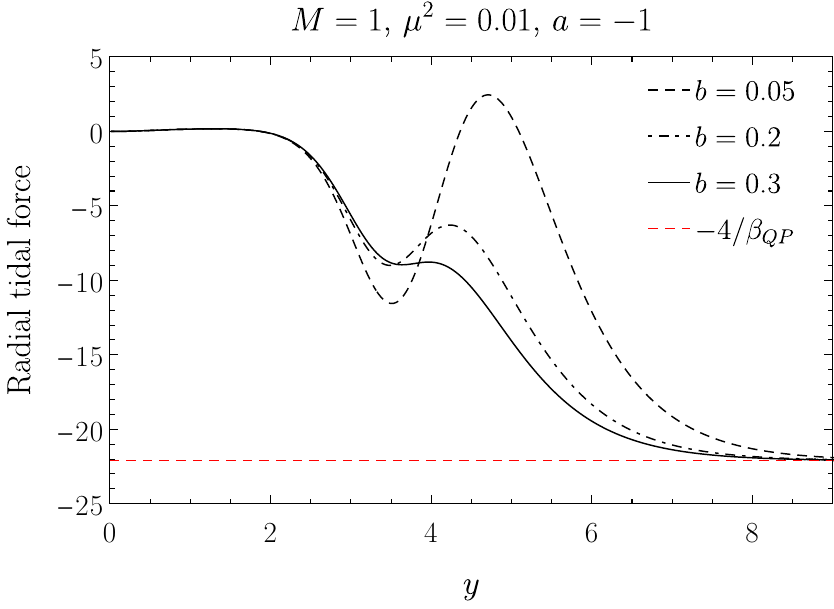}
	\caption{Radial tidal force in the QP asymmetric wormhole as a function of the radial coordinate $y$ for three values of $b$. We notice that the radial tidal force may present local maxima and minima depending on the configuration. In the asymptotic internal region, the radial tidal force goes to a finite, negative constant value. The red dashed line is the radial tidal force computed via the approximated line element~\eqref{eq:approx_metric} (cf. Eq.~\eqref{eq:rad_tid}), which correctly captures the asymptotic behavior.}
	\label{fig:rad_tidal}
\end{figure}

\begin{figure}[!h]
\center
	\includegraphics[width=\columnwidth]{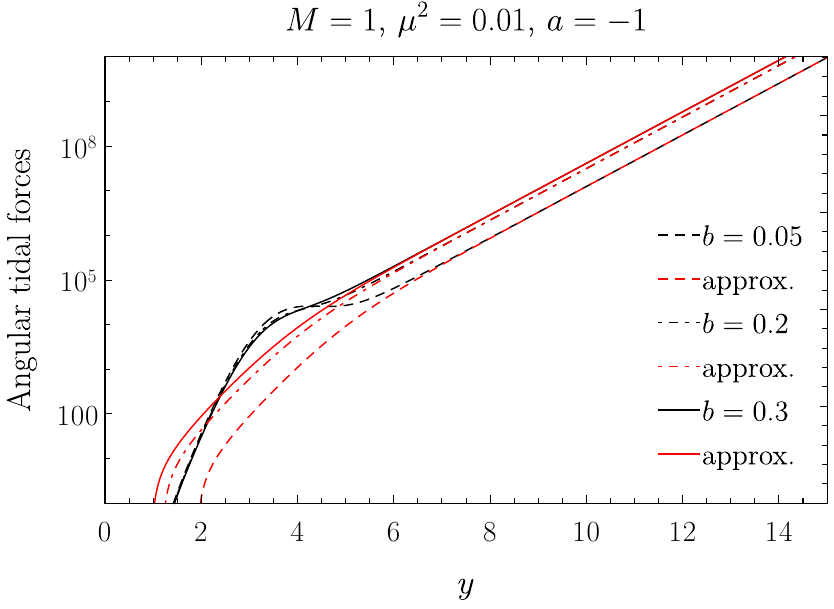}
	\caption{Angular tidal forces in the QP asymmetric wormhole as as function of the radial coordinate $y$ for three values of $b$. We notice that inside the wormhole the angular tidal forces are positive everywhere and increase without any bound as $y\to\infty$. The red lines represent the angular tidal forces computed via the approximated line element~\eqref{eq:approx_metric} (cf. Eq.~\eqref{eq:ang_tid}), which are in excellent agreement with the exact expressions even for not too large values of $y$.}
	\label{fig:ang_tidal}
\end{figure}

\begin{figure}
\includegraphics[width=\columnwidth]{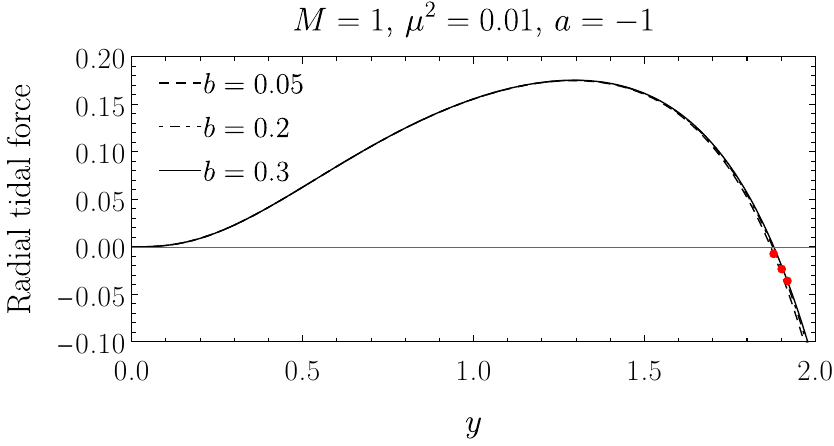}
	\includegraphics[width=\columnwidth]{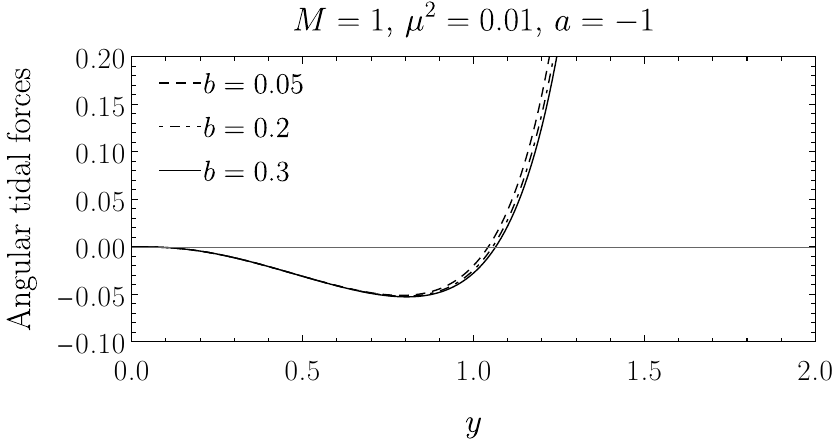}
	\caption{Radial (top) and angular (bottom) tidal forces in the outer region of the wormhole. The red dots in the top panel represent the throat of each wormhole. We notice that the radial tidal force has a local maximum, while the angular tidal forces present a local minimum. Additionally, the tidal forces may vanish before crossing the throat.}
	\label{fig:zoom_tidal}
\end{figure}

The complete equations for the tidal forces exerted by the QP wormhole are cumbersome and unwieldy, which motivates the selection of concrete model parameters to represent their behavior. In particular, we have chosen several values of the theory constant $b$ and fixed values of $M$, $\mu^2$ and $a$. In Figs.~\ref{fig:rad_tidal} and~\ref{fig:ang_tidal} we plot, respectively, the radial tidal force, $\ddot{\xi}^{\hat{1}}={K^{\hat{1}}}_{\hat{1}}\xi^{\hat{1}}$,  and the angular tidal forces, $\ddot{\xi}^{\hat{i}}={K^{\hat{i}}}_{\hat{i}}\xi^{\hat{i}}$ ($i=2,3$), as functions of $y$. We also exhibit a better view of the tidal forces in the outside region, in Fig.~\ref{fig:zoom_tidal}.
We notice that, in the outside region, the radial tidal force is small, close to zero, but always  positive, and presents a local maximum. Near the throat, but still in the outside region, the radial tidal force decreases, and may vanish and become negative before reaching the throat. In the internal region, depending on the configuration, one may notice that other maxima and minima may arise, producing a small oscillatory behavior. Far from the throat, the radial tidal force goes to an asymptotically constant value. We observe excellent agreement between the asymptotic limit of the radial tidal force and the value obtained using the approximation outlined in Sec.~\ref{sec:appx}. Particularly noteworthy is the fact that this limit is independent of $b$.
For smaller values of the theory parameter $b$, the local maximum and minimum in the internal region are more noticeable. In particular, the internal local maximum may have a positive value (cf. Fig.~\ref{fig:rad_tidal}), such that more zeroes of the radial tidal force may appear inside the throat.


The angular tidal forces, on the other hand, are negative for small values of $y$, and present a local minimum in the outside region.
They vanish just once, and it happens before reaching the throat. After that, they are always positive. By moving towards the internal region, the angular tidal forces rapidly increase without bound. A glance at Fig.~\ref{fig:ang_tidal} shows that in the asymptotic internal region the angular tidal force has an exponential growth in terms of the $y$ radial coordinate.
Again, the results for the asymptotic internal region are in excellent agreement with the approximation considered in Sec.~\ref{sec:appx} (cf. Fig.~\ref{fig:ang_tidal}). Upon inspecting Eqs.~\eqref{eq:beta_qp} and~\eqref{eq:ang_tid}, it becomes apparent that, for the QP wormhole, the divergent term of the angular tidal forces is proportional to $b^2$. Therefore, the divergent nature is more pronounced when larger values of the theory parameter $b$ are considered, as can be seen in Fig.~\ref{fig:ang_tidal}.

From the above analysis, one can summarize what happens with observers coming from the asymptotically flat region and moving toward the asymptotic internal region of the QP wormhole. Far from the throat in the asymptotically flat region ($y\approx 0$), the tidal forces are almost zero. As the observers approach the throat they experience a stretching in the radial direction and compression in the angular directions. Before crossing the throat, the tidal forces may change their signs, and the observers begin to be stretched in the angular directions and compressed in the radial direction. After crossing the throat, the radial tidal force may present an oscillatory behavior, being able to vanish again for some configurations in the internal region. Inside the throat, observers moving toward the asymptotic region face a negative radial tidal force that tends to a constant value (see Eq.~\eqref{eq:rad_tid}), and at the same time they suffer an unbounded stretching in the directions perpendicular to the radial one, caused by the divergent angular tidal forces.

\section{Final remarks}\label{sec:final}
We have investigated the tidal forces and geometric structure of a family of asymmetric wormholes engendered by self-gravitating scalar fields in two Ricci-based gravity models, namely, Palatini $f({\cal R},{\cal R}_{(\mu\nu)}{\cal R}^{(\mu\nu)})={\cal R}+a{\cal R}^2+b{\cal R}_{(\mu\nu)}{\cal R}^{(\mu\nu)}$, and the EiBI model. The study of these wormholes is motivated by a nontrivial coincidence, as these two completely different gravitational settings develop internal regions with the same functional dependence (up to model dependent constants), which can be obtained in analytical form. We investigated the geodesic structure and curvature invariants of this internal region and found that, irrespective of the behavior of the algebraic curvature scalars, which turn out to be bounded everywhere, observers and light-rays in radial motion will reach the asymptotic internal infinity in a finite affine time, which clearly indicates that the space-time is singular. 

By analyzing the tidal forces in the interior region, we found that the radial component of the tidal tensor is constant and negative in the asymptotic internal region, while the angular components of the tidal tensor are positive and grow without bound. This means that, although the curvature scalars are bounded, there are unbounded components of the curvature tensor in a parallelly propagated basis along the incomplete curves. As a result, the singularity in the interior region may be classified as a non-curvature singularity.

To delve deeper into the physical consequences of the singularity in the internal wormhole region, we studied the evolution of Jacobi fields as the singularity is approached. We  found that those fields diverge, leading to unbounded angular tidal forces. According to this, the singularity can be regarded as a deformationally strong curvature singularity. More precisely, any volume element following the incomplete geodesic paths diverges as the singularity is approached, ripping apart its structure because of the infinite tidal forces perpendicular to the radial motion.

For completeness, we also studied the tidal forces in the entire space-time of the geodesically incomplete asymmetric wormhole that arises in the quadratic Palatini model $f({\cal R},{\cal R}_{(\mu\nu)}{\cal R}^{(\mu\nu)})={\cal R}+a{\cal R}^2+b{\cal R}_{(\mu\nu)}{\cal R}^{(\mu\nu)}$. Our analysis shows that the tidal forces vanish far from the throat in the outer region, where the space-time is asymptotically flat.
When approaching the throat, both radial and angular tidal forces can change their signs. Depending on the theory parameters, other zeroes may appear in the radial tidal force inside the throat. In contrast, the angular tidal forces remain positive and exponentially growing throughout the interior region. We compared our results for the tidal forces in the asymptotic internal region with those obtained using the asymptotic metric approximation, finding excellent agreement between the two approaches.

Our study shows that the absence of unbounded algebraic curvature invariants constructed from polynomials of the Riemann tensor cannot guarantee the absence of strong singularities. Therefore, relying solely on the analysis of algebraic curvature scalars is not sufficient to assess the regularity of space-times. In alternative theories of gravity, particularly those based on the Palatini framework, the lack of correlation between curvature invariants and the geodesic structure of space-times seems to be more evident~\cite{Gonzalo:2016}. Our investigation of the geodesically incomplete asymmetric wormhole provides a concrete illustration of this lack of correlation. Examining these non-standard naked singularities may enhance our understanding of the physical significance of singularities in geometric gravity theories.
\begin{acknowledgments}
The authors would like to acknowledge 
Funda\c{c}\~ao Amaz\^onia de Amparo a Estudos e Pesquisas (FAPESPA), 
Conselho Nacional de Desenvolvimento Cient\'ifico e Tecnol\'ogico (CNPq)
 and Coordena\c{c}\~ao de Aperfei\c{c}oamento de Pessoal de N\'ivel Superior (CAPES) -- Finance Code 001, from Brazil, for partial financial support. H.L.J. and L.C. thank University of Aveiro, in Portugal; and R. M. thanks the University of Valencia, in Spain, for the kind hospitality during the completion of this work.
This work is supported by the Spanish Grant PID2020-116567GB- C21 funded by MCIN/AEI/10.13039/501100011033, and the project PROMETEO/2020/079 (Generalitat Valenciana). 
This research has further been supported by the European Union's Horizon 2020 research and innovation (RISE) programme H2020-MSCA-RISE-2017 Grant No. FunFiCO-777740 and by the European Horizon Europe staff exchange (SE) programme HORIZON-MSCA-2021-SE-01 Grant No. NewFunFiCO-101086251.
\end{acknowledgments}

\appendix
\onecolumngrid
\bigskip

\section{Zakhary and McIntosh invariants}\label{sec:zm}
Zakhary and McIntosh introduced the first complete set of algebraic curvature invariants of the Riemann tensor~\cite{Zakhary:1997}. This set corresponds to 17 scalars, being 4 Weyl invariants, 4 Ricci invariants and 9 mixed invariants, namely~\cite{Zakhary:1997,Bittencourt:2014,Overduin:2020,Kraniotis:2022}:
\begin{itemize}
\item Weyl invariants:
\end{itemize} 
\begin{align*}
I_1 & ={C_{\alpha\beta}}^{\mu\nu}{C_{\mu\nu}}^{\alpha\beta},&& I_2  ={C^\star_{\alpha\beta}}^{\mu\nu}{C_{\mu\nu}}^{\alpha\beta},\\
I_3 & ={C_{\alpha\beta}}^{\mu\nu}{C_{\mu\nu}}^{\lambda\rho}{C_{\lambda\rho}}^{\alpha\beta}, && I_4  ={C_{\alpha\beta}}^{\mu\nu}{C^\star_{\mu\nu}}^{\lambda\rho}{C_{\lambda\rho}}^{\alpha\beta} ; \numberthis
\end{align*}
\begin{itemize}
\item Ricci invariants:
\end{itemize} 
\begin{align*}
I_5 & = g_{\alpha\beta}R^{\alpha\beta},&&I_6  = R_{\alpha\beta}R^{\alpha\beta},\\
I_7 & = {R_{\alpha}}^\beta {R_{\beta}}^\mu {R_{\mu}}^\alpha,&&I_8  = {R_{\alpha}}^\beta {R_{\beta}}^\mu {R_{\mu}}^\nu {R_{\nu}}^\alpha ; \numberthis
\end{align*}
\begin{itemize}
\item Mixed invariants:
\end{itemize}
\begin{align*}
I_9 & = {C_{\alpha\beta\mu}}^{\nu}R^{\beta\mu}{R_\nu}^{\alpha},\\I_{10}&=-{C^\star_{\alpha\beta\mu}}^{\nu}R^{\beta\mu}{R_\nu}^{\alpha},\\I_{11}  &= R^{\alpha\beta}R^{\mu\nu}\left({{C_{\lambda\alpha}}^{\rho}}_{\beta}{{C_{\rho\mu}}^{\lambda}}_{\nu}-{{C^{\star}_{\lambda\alpha}}^{\rho}}_{\beta}{{C^{\star}_{\rho\mu}}^{\lambda}}_{\nu}\right)\\
I_{12}  &= -R^{\alpha\beta}R^{\mu\nu}\left({{C^{\star}_{\lambda\alpha\beta}}^{\rho}}{{C_{\rho\mu}}^{\lambda}}_{\nu}+{{C_{\lambda\alpha\beta}}^{\rho}}{{C^{\star}_{\rho\mu}}^{\lambda}}_{\nu}\right),\\
I_{13}  &= {C^{\alpha\beta}}_{\mu\nu}{S_\alpha}^{\lambda}{S_\lambda}^{\mu}{S_\beta}^{\rho}{S_\rho}^{\nu},\\
I_{14}  &= {{C^{\star}}^{\alpha\beta}}_{\mu\nu}{S_\alpha}^{\lambda}{S_\lambda}^{\mu}{S_\beta}^{\rho}{S_\rho}^{\nu},\\
I_{15}  &= \dfrac{1}{16}R^{\alpha\beta}R^{\mu\nu}\left({C_{\lambda\alpha\beta\rho}}{{C^{\lambda}}_{\mu\nu}}^{\rho}+{C^{\star}_{\lambda\alpha\beta\rho}}{{{C^{\star}}^{\lambda}}_{\mu\nu}}^{\rho}\right),\\
I_{16}  &= -\dfrac{1}{32}R^{\alpha\beta}R^{\mu\nu}\left({C_{\lambda\kappa\sigma\rho}}{{C^{\lambda}}_{\alpha\beta}}^{\rho}{{C^{\kappa}}_{\mu\nu}}^{\sigma}+{C_{\lambda\kappa\sigma\rho}}{{{C^\star}^{\lambda}}_{\alpha\beta}}^{\rho}{{{C^\star}^{\kappa}}_{\mu\nu}}^{\sigma}-{{C^\star}_{\lambda\kappa\sigma\rho}}{{{C^\star}^{\lambda}}_{\alpha\beta}}^{\rho}{{C^{\kappa}}_{\mu\nu}}^{\sigma}+{{C^\star}_{\lambda\kappa\sigma\rho}}{{C^{\lambda}}_{\alpha\beta}}^{\rho}{{{C^\star}^{\kappa}}_{\mu\nu}}^{\sigma}\right),\\
I_{17}  &= \dfrac{1}{32}R^{\alpha\beta}R^{\mu\nu}\left({{C^\star}_{\lambda\kappa\sigma\rho}}{{C^{\lambda}}_{\alpha\beta}}^{\rho}{{C^{\kappa}}_{\mu\nu}}^{\sigma}+{{C^\star}_{\lambda\kappa\sigma\rho}}{{{C^\star}^{\lambda}}_{\alpha\beta}}^{\rho}{{{C^\star}^{\kappa}}_{\mu\nu}}^{\sigma}-{C_{\lambda\kappa\sigma\rho}}{{{C^\star}^{\lambda}}_{\alpha\beta}}^{\rho}{{C^{\kappa}}_{\mu\nu}}^{\sigma}+{C_{\lambda\kappa\sigma\rho}}{{C^{\lambda}}_{\alpha\beta}}^{\rho}{{{C^\star}^{\kappa}}_{\mu\nu}}^{\sigma}\right),\\
\numberthis
\end{align*}
where ${S_{\alpha\beta}} \equiv {R_{\alpha\beta}}-(R/4)g_{\alpha\beta}$ is the trace-free Ricci tensor, $C_{\alpha\beta\mu\nu}$ is the Weyl tensor and $C^{\star}_{\alpha\beta\mu\nu}\equiv (1/2)\epsilon_{\alpha\beta\lambda\rho}{C^{\lambda\rho}}_{\mu\nu}$ is the dual of the Weyl tensor. 

The non-vanishing ZM invariants of the line element~\eqref{eq:approx_metric} are

\begin{align*}
I_1 &= \frac{4 \left(\beta -6 r^2\right)^2}{3 \beta ^2 r^4},&&I_3 = \frac{4 \left(\beta -6 r^2\right)^3}{9 \beta ^3 r^6},&&I_5 = \frac{2}{r^2}-\frac{6}{\beta },\\ 
I_6 &= \frac{36}{\beta ^2}+\frac{2}{r^4},&& I_7 = \frac{2}{r^6}-\frac{216}{\beta ^3},&&I_8 =  2 \left(\frac{648}{\beta ^4}+\frac{1}{r^8}\right),\\
I_9 &= \frac{24}{\beta ^2 r^2}-\frac{2}{3 r^6},&&I_{11} = \frac{2 \left(\beta -6 r^2\right)^2 \left(2 \beta ^2+12
   \beta  r^2+27 r^4\right)}{9 \beta ^4 r^8},&&I_{13} =-\frac{6 \left(\beta ^2-36 r^4\right)}{\beta ^4 r^6},\\
I_{15} &=\frac{\left(\beta -6 r^2\right)^2 \left(2 \beta ^2+12 \beta 
   r^2+27 r^4\right)}{72 \beta ^4 r^8},&&I_{16} = \frac{\left(\beta -6 r^2\right)^3 \left(2 \beta +3
   r^2\right) \left(2 \beta +9 r^2\right)}{432 \beta ^5
   r^{10}}.   
\numberthis
\end{align*}
\twocolumngrid
{}

\begin{thebibliography}{99}
\bibitem{Joshi:1993} P. S. Joshi and I. H. Dwivedi, Naked singularities in spherically symmetric inhomogeneous Tolman-Bondi dust cloud collapse, Phys. Rev. D \textbf{47}, 5357 (1993).
\bibitem{Mosani:2020} K. Mosani, D. Dey, and P. S. Joshi, Strong curvature naked singularities in spherically symmetric perfect fluid collapse, Phys. Rev. D \textbf{101}, 044052 (2020), [Erratum: Phys. Rev. D \textbf{107}, 069903 (2023)] 
\bibitem{Penrose:1969} R. Penrose, Gravitational Collapse: the Role of General Relativity, Nuovo Cimento Soc. Ital. Fis. \textbf{1}, 252 (1969).
\bibitem{Gyulchev:2019} G. Gyulchev, P. Nedkova, T. Vetsov, and S. Yazadjiev, Image of the Janis-Newman-Winicour naked singularity with a thin accretion disk, Phys. Rev. D \textbf{100}, 024055 (2019). 

\bibitem{Penrose:1965} R. Penrose, Gravitational collapse and space-time singularities, Phys. Rev. Lett. \textbf{14}, 57 (1965).

\bibitem{Senovilla:2014gza}
J.~M.~M.~Senovilla and D.~Garfinkle,
The 1965 Penrose singularity theorem,
Class. Quant. Grav. \textbf{32}, 124008 (2015).

\bibitem{Witten:2020} E. Witten, Light Rays, Singularities, and All That, Rev. Mod. Phys. \textbf{92}, 045004 (2020).
\bibitem{Geroch:1968} R. P. Geroch, What is a singularity in general relativity?, Ann. Physics \textbf{48}, 526 (1968).
\bibitem{Hawking:1973} S. Hawking and G. Ellis, \textit{The Large Scale Structure of Space-Time} (Cambridge Monographs on Mathematical Physics). Cambridge: Cambridge University Press. 1973.
\bibitem{Ellis:1979} G. F. R. Ellis and B. Schmidt, Classification of singular spacetimes, Gen. Relativ. Gravit. \textbf{10}, 989 (1979).
\bibitem{Hirschmann:2004} E. W. Hirschmann, Collapse of a scalar field in 2 + 1 gravity, A. Wang, and Y. Wu, Class. Quantum Grav. \textbf{21}, 1791 (2004).
\bibitem{Clarke:1975} C. J. S. Clarke, Singularities in globally hyperbolic space-time, Commun. Math. Phys. \textbf{41}, 65 (1975).
\bibitem{Janis:1968} A. I. Janis, E. T. Newman, and J. Winicour, Reality of the Schwarzschild Singularity, Phys. Rev. Lett. \textbf{20}, 878 (1968).
\bibitem{Wyman:1981} M. Wyman, Static spherically symmetric scalar fields in general relativity, Phys. Rev. D \textbf{24}, 839 (1981).

\bibitem{AORG:2017}
V.~I.~Afonso, G.~J.~Olmo, and D.~Rubiera-Garcia,
Scalar geons in Born-Infeld gravity,
JCAP \textbf{08}, 031 (2017).
\bibitem{MCO:2022} R. B. Magalh\~aes, L. C. B. Crispino, and G. J. Olmo, Compact objects in quadratic Palatini gravity generated by a free scalar field, Phys. Rev. D \textbf{105}, 064007 (2022).

\bibitem{TDE1}   S. V. Velzen, G. R. Farrar, S. Gezari, N. Morrell,  D. Zaritsky, L. Ostman, M. Smith, J. Gelfand, and A. J. Drake, Optical discovery of probable stellar tidal disruption flares. Astrophys. J. \textbf{741}, 73 (2011).

\bibitem{TDE2}  S. Gezari, D. C. Martin, B. Milliard, S. Basa, J. P. Halpern, K. Forster, P. G. Friedman, P. Morrisseyet, S.~G.~Neff, and D.~Schiminovich, Ultraviolet detection of the tidal disruption of a star by a supermassive black hole. Astrophys. J. Lett. \textbf{653}, L25 (2006).

\bibitem{TDE3} N.~Bade, S.~Komossa, and M.~Dahlem, Detection of an extremely soft X-ray outburst in the $H_{II}$-like  nucleus of NGC 5905. Astron.~Astrophys.~\textbf{309}, L35 (1996)


\bibitem{Olmo:2011} G. J. Olmo, Palatini Approach to Modified Gravity: $f(R)$ Theories and Beyond, Int. J. Mod. Phys. D \textbf{20}, 413 (2011).
%

\bibitem{Ferraris:1992}
M.~Ferraris, M.~Francaviglia, and I.~Volovich,
The Universality of vacuum Einstein equations with cosmological constant,
Class. Quant. Grav. \textbf{11}, 1505 (1994).
\bibitem{Borowiec:1996}
A.~Borowiec, M.~Ferraris, M.~Francaviglia, and I.~Volovich,
Universality of Einstein equations for the Ricci squared Lagrangians,
Class. Quant. Grav. \textbf{15}, 43 (1998).


\bibitem{Vollick:2003qp}
D.~N.~Vollick,
Palatini approach to Born-Infeld-Einstein theory and a geometric description of electrodynamics,
Phys. Rev. D \textbf{69}, 064030 (2004).
\bibitem{BeltranJimenez:2017}
J.~B. Jim{\'e}nez, L.~Heisenberg, G.~J.~Olmo, and D.~Rubiera-Garcia,
Born\textendash{}Infeld inspired modifications of gravity,
Phys. Rept. \textbf{727}, 1 (2018).
\bibitem{Afonso:2021aho}
V.~I.~Afonso, C.~Bejarano, R.~Ferraro, and G.~J.~Olmo,
Determinantal Born-Infeld coupling of gravity and electromagnetism,
Phys. Rev. D \textbf{105}, 084067 (2022).

\bibitem{Gonzalo:2022} G.~J. Olmo and D. Rubiera-Garcia, Some recent results on Ricci-based gravity theories, Int. J. Mod. Phys. D \textbf{31}, 2240012 (2022).
\bibitem{Afonso:2017} V. I. Afonso, C. Bejarano, J. Beltr{\'a}n Jim{\'e}nez, G. J. Olmo, and E. Orazi, The trivial role of torsion in projective invariant theories of gravity with non-minimally coupled matter fields, Class. Quant. Grav. \textbf{34}, 235003 (2017).

\bibitem{BeltranJimenez:2019} J. B. Jim{\'e}nez and A. Delhom, Ghosts in metric-affine higher order curvature gravity, Eur. Phys. J. C \textbf{79}, 656 (2019).
\bibitem{BeltranJimenez:2020} J. B. Jim{\'e}nez and A. Delhom, Instabilities in metric-affine theories of gravity with higher order curvature terms, Eur. Phys. J. C \textbf{80}, 585 (2020).

\bibitem{AORG:2018} V. I. Afonso, G. J. Olmo, and D. Rubiera-Garcia, Mapping Ricci-based theories of gravity into general relativity, Phys. Rev. D \textbf{97}, 021503(R) (2018).
\bibitem{AOORG:2018}
V.~I.~Afonso, G.~J.~Olmo, E.~Orazi, and D.~Rubiera-Garcia, Correspondence between modified gravity and general relativity with scalar fields, Phys. Rev. D \textbf{99}, 044040 (2019).
\bibitem{AOORG:2019}
V.~I.~Afonso, G.~J.~Olmo, E.~Orazi, and D.~Rubiera-Garcia, New scalar compact objects in Ricci-based gravity theories,
JCAP \textbf{12}, 044 (2019).

\bibitem{OORG:2014} S. D. Odintsov, G. J. Olmo, and D. Rubiera-Garcia, Born-Infeld gravity and its functional extensions, Phys. Rev. D \textbf{90}, 044003 (2014).
\bibitem{ORGSP:2016} G. J. Olmo, D. Rubiera-Garcia, and A. Sanchez-Puente, Classical resolution of black hole singularities via wormholes, Eur. Phys. J. C \textbf{76} 143 (2016).
\bibitem{Gonzalo:2016} G. J. Olmo, D. Rubiera-Garcia, and A. Sanchez-Puente, Impact of curvature divergences on physical observers in a wormhole space-time with horizons, Class.~Quant.~Grav. \textbf{33}, 115007 (2016). 


\bibitem{Zakhary:1997} E. Zakhary and C. B. G. McIntosh, A Complete Set of Riemann Invariants, Gen. Relativ. Gravit. \textbf{29}, 539 (1997).
%


\bibitem{Clarke:1977} C. J. S. Clarke and B. G. Schmidt, Singularities: The state of the art. Gen. Relativ. Gravit. \textbf{8}, 129 (1977).
\bibitem{Vickers:1987} J. A. G. Vickers,  Quasi-regular singularities and cosmic strings, Class. Quantum Grav. \textbf{4} 1 (1987).
\bibitem{Konkowski:1985} D. A. Konkowski, T. M. Helliwell, and L. C. Shepley, Cosmologies with quasiregular singularities. I. Spacetimes and test waves, Phys. Rev. D \textbf{31}, 1178 (1985).
%


\bibitem{Pirani_rep} F.~A.~E. Pirani, Republication of: On the physical significance of the Riemann tensor. Gen. Relativ. Gravit. \textbf{41}, 1215 (2009).

\bibitem{dinverno} R. D’Inverno, \textit{Introducing Einstein’s Relativity}. Oxford: Clarendon Press. 1992.


\bibitem{Crispino:2016} L.  C.  B.  Crispino,  A.  Higuchi,  L.  A.  Oliveira, and E. S. de Oliveira , Tidal forces in Reissner–Nordström spacetimes, Eur. Phys. J. C \textbf{76}, 168 (2016).
\bibitem{Goel:2015} A. Goel, R. Maity, P. Roy, and T. Sarkar, Tidal forces in naked singularity backgrounds, Phys. Rev. D \textbf{91}, 104029 (2015).
%

\bibitem{Tipler:1977a} F. J. Tipler, On the nature of singularities in general relativity, Phys. Rev. D \textbf{15}, 942 (1977).
\bibitem{Tipler:1977b} F. J. Tipler, Singularities in conformally flat spacetimes, Phys. Lett. A \textbf{64}, 8 (1977).
\bibitem{Clarke:1985} C. J. S. Clarke and A. Kr\'olak, Conditions for the occurence of strong curvature singularities, J. Geom. Phys. \textbf{2}, 127 (1985). 
%

\bibitem{nolan} B. C. Nolan, Strengths of singularities in spherical symmetry, Phys. Rev. D \textbf{60}, 024014 (1999).
\bibitem{ori} A. Ori, Strength of curvature singularities, Phys. Rev. D \textbf{61}, 064016 (2000).

\bibitem{Bittencourt:2014} E. Bittencourt, J. M. Salim, and G. B. Santos, Magnetic fields and the Weyl tensor in the early universe. Gen. Relativ. Gravit. \textbf{46}, 1790 (2014).
\bibitem{Overduin:2020} J. Overduin, M. Coplan, K. Wilcomb, and R. C. Henry, Curvature invariants for charged and rotating black holes, Universe \textbf{6}, 22 (2020).
\bibitem{Kraniotis:2022} G. V. Kraniotis, Curvature invariants for accelerating Kerr–Newman black holes in (anti-)de Sitter spacetime, Class. Quantum Grav. \textbf{39}, 145002 (2022).






 

%

%
\end{thebibliography}
\end{document}